\documentclass[letterpaper]{article}
\usepackage[preprint]{aaai2027}
\usepackage[hyphens]{url}
\usepackage{graphicx}
\usepackage{natbib}
\usepackage{caption}
\usepackage{algorithm}
\usepackage{algorithmic}
\usepackage{amsmath,amssymb,multirow}
\usepackage{newfloat,listings}
\usepackage{booktabs,tabularx,array}
\usepackage[table]{xcolor}
\DeclareCaptionStyle{ruled}{labelfont=normalfont,labelsep=colon,strut=off}
\floatstyle{ruled}
\newfloat{listing}{tb}{lst}{}
\floatname{listing}{Listing}
\definecolor{TableHeader}{HTML}{DCEAF7}
\definecolor{TableStripe}{HTML}{F5F6F9}
\definecolor{TableOurs}{HTML}{E6D4FA}
\definecolor{TableGreen}{HTML}{008A67}
\definecolor{TableOrange}{HTML}{C77700}
\newcommand{\tblcheck}{\textcolor{TableGreen}{\checkmark}}
\newcommand{\tblpartial}{\textcolor{TableOrange}{\ensuremath{\sim}}}
\newcommand{\tbldash}{\textcolor{gray!65}{--}}
\newcolumntype{L}[1]{>{\raggedright\arraybackslash}p{#1}}
\newcolumntype{C}[1]{>{\centering\arraybackslash}p{#1}}
\newcolumntype{Y}{>{\raggedright\arraybackslash}X}
\title{WEAVE: Weaving Real-World Code Evolution into Dependency-Structured Tasks for Benchmarking and Training Code Agents}
\author{
Weihao Li\textsuperscript{\rm 1}\equalcontrib,
Zihang Xu\textsuperscript{\rm 2}\equalcontrib,
Zhongqiu Wang\textsuperscript{\rm 1}\equalcontrib,
Songwei Yu\textsuperscript{\rm 4},
Wenqiang Zhu\textsuperscript{\rm 1},
Yuechen Luo\textsuperscript{\rm 3}\\
Jingxuan Xu\textsuperscript{\rm 4},
Yanan Wu\textsuperscript{\rm 1},
Haoyang Huang\textsuperscript{\rm 1},
Ken Deng\textsuperscript{\rm 1}\corresponding
}
\affiliations{
\textsuperscript{\rm 1}Independent Researchers,
\textsuperscript{\rm 2}Shanghai Artificial Intelligence Laboratory\\
\textsuperscript{\rm 3}Tsinghua University,
\textsuperscript{\rm 4}National University of Singapore\\
\texttt{bairidreamer.cn@gmail.com}
}

\begin{document}
\maketitle
\begin{abstract}
Real-world software engineering rarely consists of isolated changes.
Features, migrations, and systematic fixes often unfold across multiple
pull requests (PRs), requiring coordinated modifications across a
repository. Yet most executable repository-level benchmarks retain a
single issue or PR as the task unit, while existing multi-PR settings rely
on coarse aggregation and structure-agnostic evaluation. This mismatch
obscures dependencies among changes and limits progress measurement and
learning signals for long-horizon code agents. We introduce \textsc{WEAVE}, a framework that transforms verified
PR histories into executable, dependency-structured tasks.
\textsc{WEAVE} leverages repository-calibrated typed change loci to organize related PRs into bounded DAGs and instantiate each as a unified executable
task. We further introduce Graph Credit, a deterministic, test-grounded metric that measures dependency-consistent partial progress and supports dense, verifiable rewards for reinforcement learning.
Experiments on task construction, agent evaluation, supervised
fine-tuning, and reinforcement learning show that \textsc{WEAVE} yields
coherent executable tasks, enables more faithful progress measurement than
flat scoring, and provides effective supervision and verifiable rewards
for training code agents.
\end{abstract}
\section{Introduction}
\label{sec:introduction}

\begin{table*}[t]
    \centering
    \caption{
    Comparison of executable software-engineering benchmarks
    \textbf{Struct.} denotes task structure retained for evaluation;
    \textbf{Exec.} indicates an executable oracle; and
    \textbf{Train} indicates available executable training data.
    \tblpartial{} denotes partial support.
    }
    \label{tab:benchmark_comparison}

    \small
    \setlength{\tabcolsep}{3.3pt}
    \renewcommand{\arraystretch}{1.14}

    \begin{tabularx}{\textwidth}{
        L{3.0cm}
        L{2.3cm}
        L{2.9cm}
        C{1.80cm}
        C{1.30cm}
        C{1.90cm}
        C{1.30cm}
        C{1.40cm}
    }
        \toprule

        \textbf{Benchmark}
        & \textbf{Unit}
        & \textbf{Construction}
        & \textbf{Struct.}
        & \textbf{Exec.}
        & \textbf{Credit}
        & \textbf{Train}
        & \textbf{Lang.}
        \\
        \midrule

        SWE-bench Verified
        & Issue--PR
        & Issue linked
        & Atomic
        & \tblcheck
        & Binary
        & \tbldash
        & Py
        \\

        Multi-SWE-bench
        & Issue--PR
        & Issue linked
        & Atomic
        & \tblcheck
        & Binary
        & \tblcheck
        & Multi
        \\

        SWE-Gym
        & Issue--PR
        & Real issues
        & Atomic
        & \tblcheck
        & Binary
        & \tblcheck
        & Py
        \\

        \midrule

        SWE-EVO
        & Release delta
        & Release based
        & Flat
        & \tblcheck
        & Flat F2P
        & \tbldash
        & Py
        \\

        SWE-CI
        & Commit span
        & Evolution span
        & Sequence
        & \tblcheck
        & EvoScore
        & \tbldash
        & Py
        \\

        SWE-Milestone
        & Milestone
        & Commit + LLM
        & DAG
        & \tblcheck
        & Milestone
        & \tbldash
        & Multi
        \\

        daVinci-Agency
        & PR chain
        & Explicit references
        & Chain
        & \tblpartial
        & \tbldash
        & \tblcheck
        & Multi
        \\

        \midrule

        \textbf{\textsc{WEAVE} (Ours)}
        & \textbf{Coupled PRs}
        & \textbf{Typed-locus coupling}
        & \textbf{DAG}
        & \tblcheck
        & \textbf{Graph Credit}
        & \tblcheck
        & \textbf{Multi}
        \\

        \bottomrule
    \end{tabularx}
\end{table*}

\begin{figure}[t]
   \centering
    \includegraphics[width=0.47\textwidth]{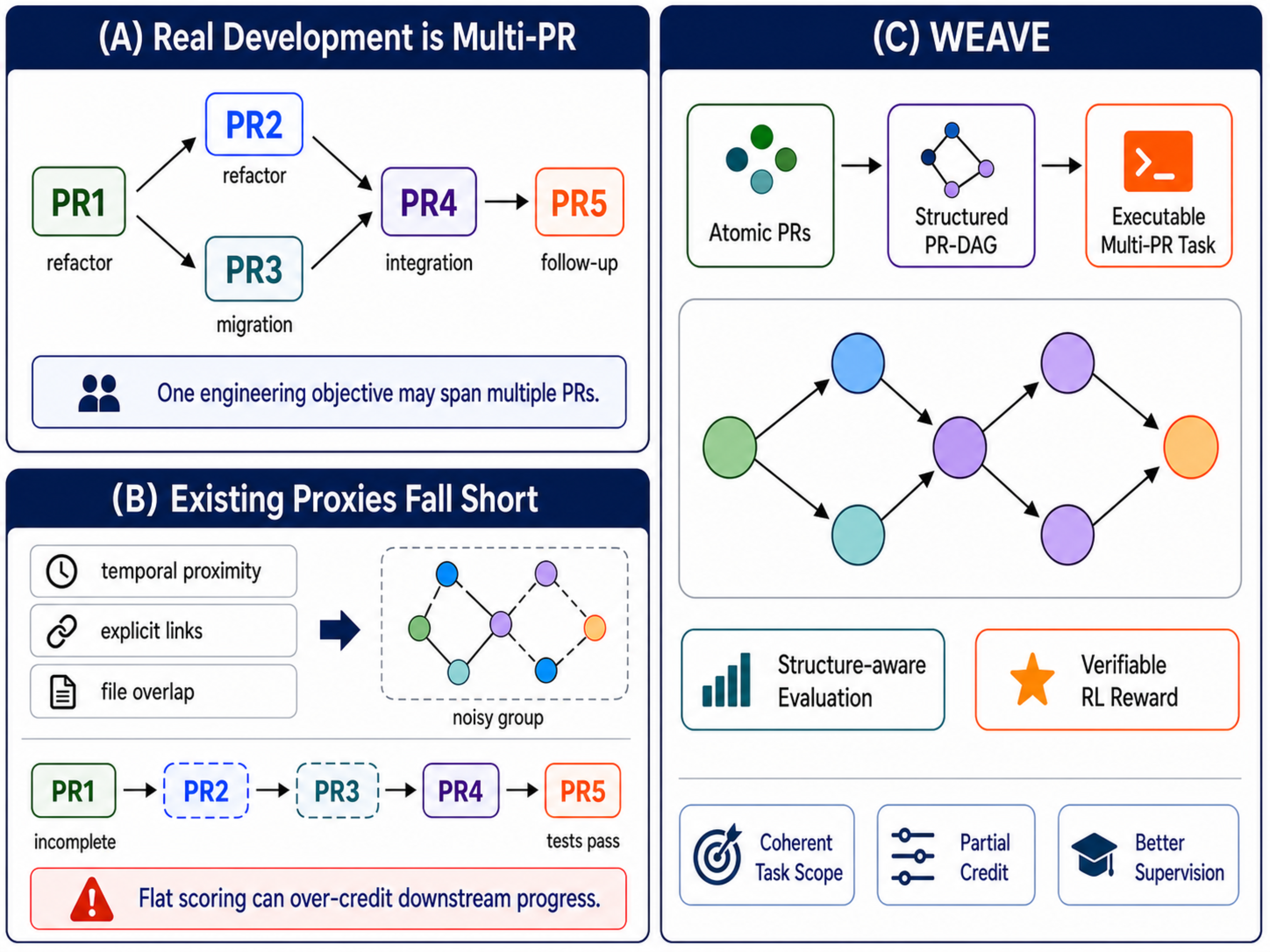}
    \caption{Real-world objectives span dependent PRs, but existing proxies obscure task structure and credit assignment. WEAVE constructs executable dependency-structured tasks for verifiable evaluation and rewards.}

    \label{fig:weave_motivation}
\end{figure}

Executable repository-level benchmarks have increasingly become a central testbed for
software engineering agents~\citep{yang2025codeintelligencesurvey}. SWE-bench and its variants pair a
natural-language issue with a repository state and an executable test
oracle, enabling standardized, verifiable evaluation of generated
patches~\citep{jimenez2023swebench,
zan2025multiswebench,deng2025swebenchpro}. Most benchmarks,
however, retain a single issue or pull request (PR) as the basic task
unit. This abstraction captures isolated bug fixes and localized features,
but only a restricted portion of real software development. In practice,
a single feature, migration, dependency upgrade, or systematic fix often evolves across
multiple PRs. These changes may be sequential, branch in parallel, or
converge through later integration. The missing abstraction is therefore
not merely longer context or a larger patch, but the \emph{dependency
structure} connecting multiple changes into a coherent engineering
objective. Figure~\ref{fig:weave_motivation} summarizes this mismatch
between dependency-structured development, proxy-based task construction,
and structure-agnostic credit assignment.

Recent work has moved beyond isolated PRs by grouping changes through
release windows, temporal spans, explicit cross-references, or object-level
dependencies~\citep{thai2025sweevo,chen2026sweci,
jiang2026davinci,zhou2026featurebench}. These efforts advance long-horizon
software engineering, yet two coupled challenges remain. First, these aggregation signals are imperfect proxies
for task coherence: temporal proximity can merge unrelated changes,
explicit references are sparse, and coarse overlap is often dominated by
frequently modified utilities, configuration files, or dependency
manifests. Moreover, a related change set does not itself define a
reproducible task with a valid base state, task specification, and
executable oracle. Second, current evaluation poorly captures structured
partial progress. Binary resolution ignores progress on unsolved tasks,
while flat test- or PR-level aggregation overlooks dependencies and may
over-credit downstream behavior built on incomplete upstream changes. The
same issue weakens reinforcement learning: binary rewards are sparse,
whereas patch-similarity and learned rewards may not align with executable
task semantics~\citep{hybridrl2025,wei2025swerl,verpo2026,
shum2025swerm}. Addressing these limitations requires a unified,
code-grounded representation for executable task construction,
dependency-aware evaluation, and verifiable learning signals.

We introduce \textsc{WEAVE}, a framework for weaving real-world code
evolution into executable, dependency-structured tasks for benchmarking
and training code agents. As illustrated in
Figure~\ref{fig:weave_motivation}, \textsc{WEAVE} starts from verified
atomic PRs, identifies code-grounded relations through typed change loci
calibrated by repository-local frequencies, and organizes related changes
into bounded dependency DAGs. Each DAG is instantiated as a reproducible
repository-level task with a unified specification and executable oracle,
while remaining evaluator-side metadata unavailable to the agent. Beyond
task construction, the DAG provides a common interface for measuring and
learning from structured progress. We introduce Graph Credit, a
deterministic and test-grounded metric that complements binary resolution
by rewarding dependency-consistent partial completion. The same structure
supports dense, verifiable rewards for reinforcement learning, while the
constructed corpus provides long-horizon trajectories for supervised
fine-tuning. \textsc{WEAVE} thereby unifies executable task construction,
structure-aware evaluation, and agent training over realistic multi-PR
software evolution.

To assess both the benchmarking and training utility of \textsc{WEAVE},
we conduct a four-part empirical study covering task construction, agent
evaluation, supervised fine-tuning, and reinforcement learning.
\textsc{WEAVE} yields 3200 training tasks and
300 held-out benchmark tasks spanning
six programming languages. Its code-grounded aggregation
produces more coherent task groupings and a higher executable-task yield
than temporal and coarse-overlap baselines. Evaluations of
eight software engineering agents further show that Graph Credit
exposes structurally inconsistent credit obscured by flat scoring. Under
matched supervised-token budgets, \textsc{WEAVE}-based supervision
improves downstream agent performance, while graph-structured rewards
outperform binary and flat-reward baselines. Together, these results
establish dependency-structured tasks as a principled foundation for both
evaluating and improving code agents.

\paragraph{Contributions.}
Our contributions are threefold:
\begin{itemize}
    \item \textbf{Executable dependency-structured task construction.}
    We introduce \textsc{WEAVE}, a framework that constructs bounded
    dependency DAGs from verified PR histories using
    repository-calibrated typed change loci, and instantiates each graph
    as a unified executable task.

    \item \textbf{Structure-aware evaluation and verifiable credit.}
    We introduce Graph Credit, a deterministic, test-grounded metric for
    evaluating dependency-consistent partial progress. By conditioning
    credit on upstream resolution, it complements binary and flat scoring
    while providing a dense, verifiable signal for agent optimization.

    \item \textbf{Empirical validation for benchmarking and training.} Extensive experiments show that \textsc{WEAVE} yields coherent executable tasks, captures progress missed by structure-agnostic evaluation, and
    provides effective supervision and verifiable rewards for training code agents.
\end{itemize}

\section{Related Work}
\label{sec:related_work}

\begin{figure*}[t]
    \centering
    \includegraphics[width=\textwidth]{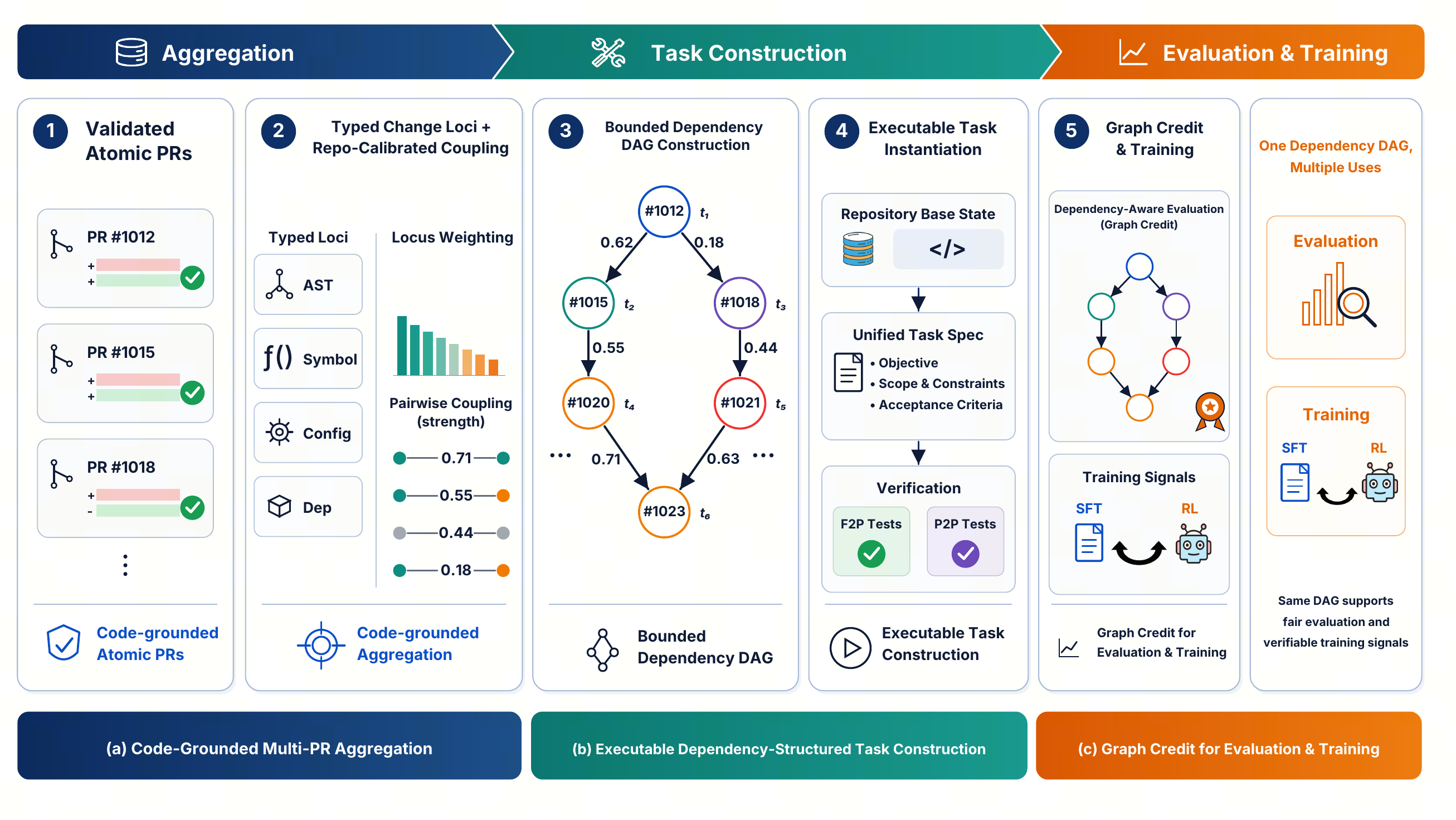}
    \caption{
    Overview of \textsc{WEAVE}. Validated PRs are represented by typed change
    loci, coupled into bounded dependency DAGs, and instantiated as executable
    tasks with unified specifications and node-attributed test oracles. The
    resulting graphs support structure-aware evaluation and reinforcement
    learning.
    }
    \label{fig:weave_pipeline}
\end{figure*}

Table~\ref{tab:benchmark_comparison} compares representative executable
software-engineering benchmarks by task unit, construction, retained
structure, evaluation granularity, and training support.

\subsection{Single-PR Software Engineering Benchmarks}
\label{sec:related_single_pr}

SWE-bench established executable repository-level evaluation by pairing
an issue and a pre-change repository state with F2P/P2P tests
\citep{jimenez2023swebench}. Subsequent benchmarks improve instance
quality or expand the same atomic issue--PR formulation across visual
software, programming languages, recent repository states, and more
challenging enterprise-scale changes
\citep{jimenez2023swebench,yang2024swebenchmm,zan2025multiswebench,
zhang2025swebenchlive,deng2025swebenchpro}. Executable workbenches such
as SWE-Gym, R2E-Gym, SWE-Smith, and SWE-rebench V2 further scale this
primitive for trajectory generation and model training
\citep{pan2024swegym,jain2025r2egym,
yang2025swesmithscalingdatasoftware,badertdinov2026swerebench}.
Together with recent contamination analyses, these works establish the foundations of
verifiable SWE-agent evaluation and training, while generally retaining
an individual issue, commit, or PR as the basic task unit.

Complementary evaluations cover repository-level code completion
\citep{deng2024r2c2coder,yang2024execrepobench}, broad agentic coding
capabilities~\citep{xu2025swecompass}, multimodal web development
\citep{lei2026webcompass}, engineering-constraint compliance
\citep{zhu2026anchor}, and code critique~\citep{zhang2025codecriticbench}.
These axes broaden capability assessment but address different questions
from dependency-consistent progress across coupled PRs.

\subsection{Multi-PR Software Engineering Benchmarks}
\label{sec:related_multi_pr}

Recent work extends software-engineering evaluation to broader units by
aggregating changes across releases, features, or development roadmaps
\citep{thai2025sweevo,zhou2026featurebench,xu2026roadmapbench}, or by
preserving repository evolution through temporal PR ranges, commit spans,
release chains, and chronologically linked fixes
\citep{shastry2026swesteps,chen2026sweci,lam2026swechain,
jin2026chainswe}. Other approaches reconstruct milestone DAGs or mine
explicitly referenced PR chains
\citep{deng2026swemilestone,jiang2026davinci}. These benchmarks commonly
measure progress through flat test completion, stage- or component-level
scores, or milestone resolution. Related work has also explored denser
signals based on patch similarity, weighted tests, learned reward models,
process feedback, enhanced test oracles, and multidimensional evaluation
\citep{wei2025swerl,verpo2026,shum2025swerm,psgpo2024,
shen2025utboost,hybridrl2025,sweperf2025,swelancer2025,xu2026flare}, while graph
structures and cross-language code representations have been used to
organize agent workflows or support structural program analysis
\citep{chen2024coder,song2024tsed,cast2025,codebasememory2026}.
Complementary work explores trajectory-level failure diagnosis
\citep{li2026codetracer}, vision-grounded feedback for web-coding
reinforcement learning~\citep{li2025relook}, and multi-stage training
and efficient adaptive reasoning~\citep{zhan2025katcoder,zhan2025katv1,deng2025hipo}.
\textsc{WEAVE} complements these directions by constructing executable
PR-DAG tasks from repository-calibrated code coupling and deriving
graph-conditioned partial credit from node-attributed tests, which can
also serve as a dense, verifiable outcome signal during optimization.

\section{Dependency-Structured Task Construction}
\label{sec:construction}

\subsection{Overview and Task Definition}
\label{sec:overview}

\textsc{WEAVE} constructs executable dependency-structured tasks from
validated atomic PR instances. As illustrated in
Figure~\ref{fig:weave_pipeline}, the construction contains two stages.
First, \textsc{WEAVE} represents each PR by the program entities it
modifies and aggregates strongly coupled PRs into bounded dependency
DAGs. Second, each candidate graph is instantiated
as a single-environment task with a unified specification, executable
tests, and node-level test attribution. The resulting graph is retained
as evaluator-side metadata and is subsequently used for the
graph-conditioned credit and reinforcement-learning methods in
Section~\ref{sec:credit_rl}.

A final \textsc{WEAVE} instance is defined as
\begin{equation}
    \mathcal{I}
    =
    \left(
        S_{\mathrm{base}},
        q,
        G,
        \mathcal{T}_{\mathrm{F2P}},
        \mathcal{T}_{\mathrm{P2P}},
        A
    \right),
\end{equation}
where $S_{\mathrm{base}}$ is the repository state presented to the agent,
$q$ is a unified task specification, $G=(V,E)$ is the dependency DAG, $\mathcal{T}_{\mathrm{F2P}}$ and
$\mathcal{T}_{\mathrm{P2P}}$ are the functional and regression test
sets, and $A$ denotes test-to-node attribution. The agent observes only
$S_{\mathrm{base}}$ and $q$; individual PR descriptions, identifiers,
historical decomposition, and the graph itself are not exposed.

\subsection{Repository-Calibrated DAG Construction}
\label{sec:dag_construction}

Temporal proximity and file-level overlap are insufficient for reliable
dependency recovery: nearby PRs may be unrelated, while frequently
modified files may connect otherwise independent changes. \textsc{WEAVE}
therefore represents PRs at the level of typed change loci and calibrates
their shared evidence against repository-specific modification patterns.

\paragraph{Typed change loci.}
For each PR $p$, we extract a set of typed change loci
\begin{equation}
    \mathcal{L}(p)
    =
    \mathcal{L}_{\mathrm{ast}}(p)
    \cup
    \mathcal{L}_{\mathrm{sym}}(p)
    \cup
    \mathcal{L}_{\mathrm{cfg}}(p)
    \cup
    \mathcal{L}_{\mathrm{decl}}(p)
    \cup
    \mathcal{L}_{\mathrm{dep}}(p).
\end{equation}
These loci capture AST declarations, lightweight symbols, structured
configuration or schema entries, and dependency keys. Thus, a PR is
represented by the concrete program entities and structured entries it
changes, rather than by the files it touches. Type-specific extraction
prevents configuration files, manifests, and lockfiles from being treated
as single shared locations. Test files, generated or vendored code, and
binary diffs are excluded from aggregation. Detailed extraction rules are
provided in Appendix.

\paragraph{Repository-calibrated coupling.}
The evidential value of a locus depends on how frequently it is modified
within its repository. Let $\mathcal{P}_R$ be the candidate PR set in
repository $R$, $N_R=|\mathcal{P}_R|$, and

\(\operatorname{df}_R(\ell)
=
|\{p\in\mathcal{P}_R:\ell\in\mathcal{L}(p)\}|\)
be the number of PRs touching locus $\ell$. We assign $\ell$ the
repository-local inverse-frequency weight
\begin{equation}
    w_R(\ell)
    =
    \log
    \left(
        \frac{N_R+1}
             {\operatorname{df}_R(\ell)+1}
    \right)
    +1.
\end{equation}
The coupling score between PRs $p_i$ and $p_j$ is then defined as
\begin{equation}
    s_R(p_i,p_j)
    =
    \frac{
        \sum_{\ell\in
        \mathcal{L}(p_i)\cap\mathcal{L}(p_j)}
        w_R(\ell)
    }{
        \sum_{\ell\in
        \mathcal{L}(p_i)\cup\mathcal{L}(p_j)}
        w_R(\ell)
    }.
\end{equation}
This weighted overlap suppresses repository-wide hubs and emphasizes
specific, infrequently modified loci that provide stronger evidence of
continuous engineering work.

\paragraph{Edge filtering and bounded DAG assembly.}
Candidate relations are constructed only within the same repository:
\begingroup\small
\begin{equation}
    E_R^{\mathrm{cand}}
    =
    \left\{
        (p_i,p_j)
        \;\middle|\;
        t_i<t_j,\;
        t_j-t_i\leq\Delta,\;
        s_R(p_i,p_j)\geq\theta
    \right\},
\end{equation}
\endgroup
where $\Delta$ is the search radius and $\theta$ the minimum coupling threshold. Merge time only constrains search and orients edges; temporal proximity does not indicate coupling.

We remove relations induced by repeated or reversed changes, including reverts, backports, release replays, and near-duplicate patches. The remaining edges define repository-level connected components. Components of at most $K$ nodes are retained; larger ones are reduced to high-scoring connected subgraphs. Preserving all surviving edges among selected nodes yields bounded dependency DAGs with linear, fork, join, and mixed structures. Merge-time linearization is used only as a baseline. Appendix details further filtering and graph assembly.

\subsection{Executable Task Instantiation}
\label{sec:workbench}

A dependency DAG defines the scope of a candidate task but does not by itself
provide an executable benchmark instance. \textsc{WEAVE} therefore
constructs a unified repository state, task specification, and test oracle
for each retained graph.

\paragraph{Executable task instantiation.}
For each candidate graph \(G\), we start from a repository state \(S_{\mathrm{gold}}\) containing all constituent PRs and revert their changes to construct the counterfactual base state \(S_{\mathrm{base}}\), while preserving changes outside \(G\). We then consolidate the PR objectives into a single task specification \(q\), removing PR identifiers, commit metadata, and historical implementation details. The agent solves the task in a single repository state and returns one patch, without reproducing the original PR sequence.

The combined executable oracle must satisfy
\begin{equation}
    \mathcal{T}_{\mathrm{F2P}}(S_{\mathrm{base}})
    =
    \mathrm{fail},
    \qquad
    \mathcal{T}_{\mathrm{F2P}}(S_{\mathrm{gold}})
    =
    \mathrm{pass},
\end{equation}
while preserving the required
$\mathcal{T}_{\mathrm{P2P}}$ behavior. Candidates are discarded when
repository reconstruction, gold-patch application, or executable
validation fails.

\paragraph{Node attribution and corpus construction.}
To support graph-conditioned credit, each valid F2P test is assigned to a PR node when high-confidence attribution is possible:
\begin{equation}
    A:
    \mathcal{T}_{\mathrm{F2P}}
    \rightarrow
    V
    \cup
    \{
        \mathrm{ambiguous},
        \mathrm{unattributed}
    \}.
\end{equation}
Attributed tests enable node-level scoring, while all valid tests determine
full-task correctness. We retain only executable instances with adequate
attribution coverage and no recoverable target commits or repository history.
Appendix details reconstruction and attribution.

\begin{figure}[t]
    \centering
    \includegraphics[width=0.47\textwidth]{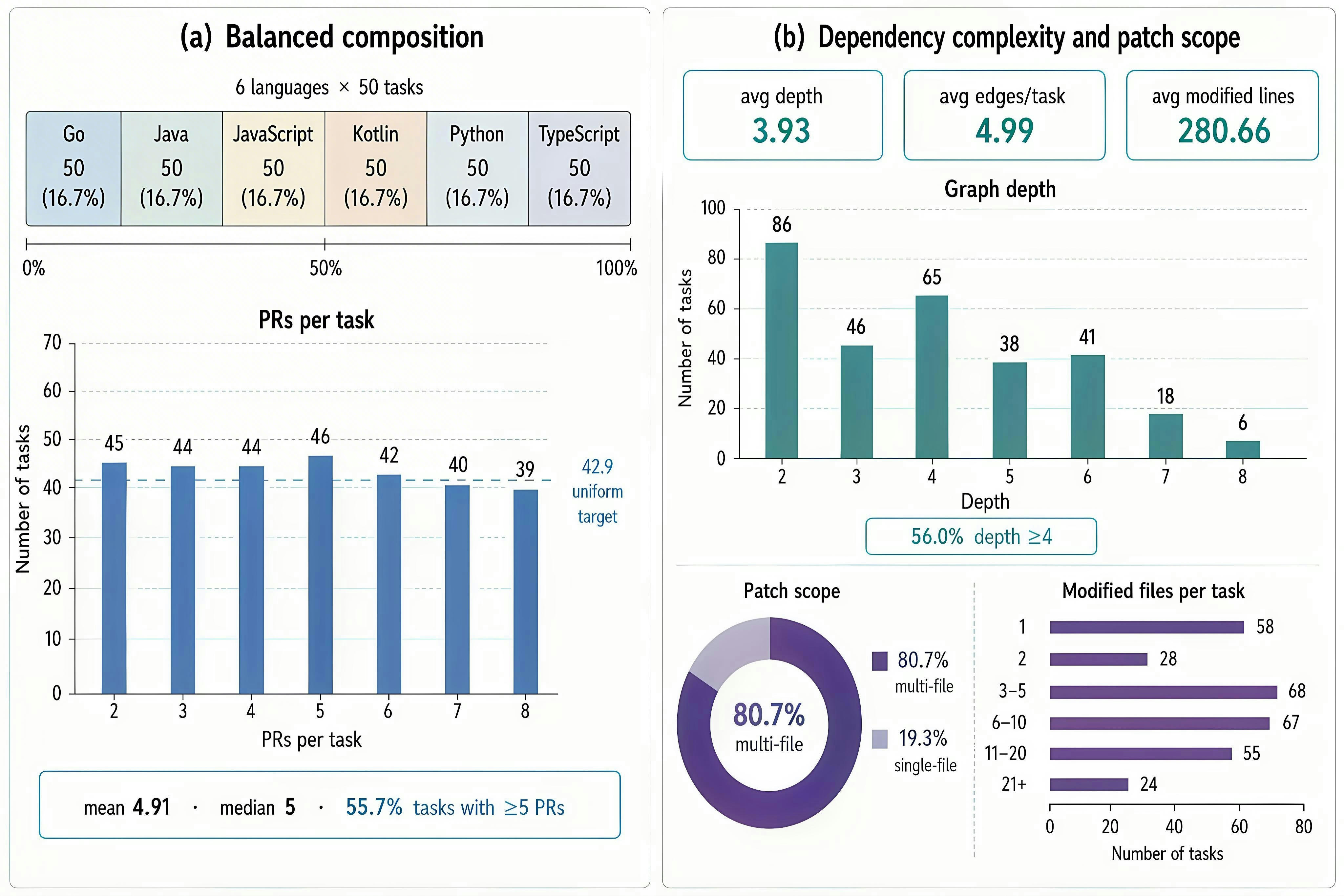}
    \caption{
        Composition and structural characteristics of \textsc{WEAVE-Bench}.
        \textbf{(a)} Language balance and PR scale.
        \textbf{(b)} Dependency-DAG complexity and patch scope.
    }
    \label{fig:weave_bench_statistics}
\end{figure}

Validated instances are partitioned into \textsc{WEAVE-Train} and the held-out
\textsc{WEAVE-Bench}. We enforce temporal separation and eliminate overlapping
target PRs, duplicate or near-duplicate patches, and other direct solution
leakage across splits. As shown in
Figure~\ref{fig:weave_bench_statistics}, \textsc{WEAVE-Bench} balances language
coverage while spanning diverse task scales, dependency structures, and patch
scopes, including non-linear PR dependencies and coordinated multi-file
changes.

\section{Graph Credit}
\label{sec:credit_rl}

Binary resolution provides a definitive measure of full-task correctness
but cannot distinguish progress among unsolved tasks. Flat test scoring
offers finer granularity but ignores dependencies: a downstream node may
pass its local tests while an upstream prerequisite remains unresolved.
We call the resulting structurally unsupported score \emph{phantom
credit}. Graph Credit addresses this limitation by conditioning
node-level completion on the task DAG.

\subsection{Dependency-Conditioned Scoring}
\label{sec:graph_credit}

For each evaluable node $v$, let $\mathcal{T}_v$ denote its confidently
attributed F2P tests. Its local completion is
\begin{equation}
    \ell_v
    =
    \frac{1}{|\mathcal{T}_v|}
    \sum_{t\in\mathcal{T}_v}
    \mathbb{I}[t\ \mathrm{passes}].
\end{equation}
Let
\(
V_A=\{v\in V:|\mathcal{T}_v|>0\}
\)
and let $G_A=(V_A,E_A)$ be the induced scoring graph. Ambiguous and
unattributed tests remain part of the full-task oracle but are excluded
from node-level scoring.

To enforce dependency consistency, we recursively define the effective
credit of node $v$ as
\begin{equation}
    c_v
    =
    \begin{cases}
        \ell_v,
        & \operatorname{pa}_{G_A}(v)=\varnothing,\\[3pt]
        \min\!\left\{
            \ell_v,\,
            \displaystyle
            \min_{u\in\operatorname{pa}_{G_A}(v)} c_u
        \right\},
        & \text{otherwise}.
    \end{cases}
\end{equation}
Because $G_A$ is acyclic, this recursion is well defined. Graph Credit is
the mean effective credit over evaluable nodes:
\begin{equation}
    \operatorname{GraphCredit}(G_A)
    =
    \frac{1}{|V_A|}
    \sum_{v\in V_A} c_v.
\end{equation}

The recursive minimum caps each node’s credit by its own completion and that of its upstream prerequisites. Thus, downstream test passes cannot receive more credit than the least-complete prerequisite on their dependency paths. Graph Credit relies solely on executable outcomes and evaluator-side graph metadata; it neither imposes the historical PR order nor uses patch similarity or learned judges.

\subsection{Evaluation and Agent Optimization}
\label{sec:evaluation_optimization}

\subsubsection{Evaluation Protocol}
\label{sec:evaluation_protocol}

We report Graph Credit alongside two complementary metrics.
\textsc{Resolved} measures full-task correctness and remains the primary
benchmark metric, while Flat-F2P measures the fraction of functional
tests passed. Graph Credit instead captures dependency-consistent partial
progress, providing a structural diagnostic for unsolved tasks.
Alternative graph operators, linearized scoring, and randomized-graph
controls are examined in RQ2.

\subsubsection{Reinforcement-Learning Reward}
\label{sec:rl_reward}

Graph Credit also provides a structural signal for reinforcement
learning. We combine flat functional progress, dependency-conditioned
credit, full-task resolution, and regression preservation:
\begin{equation}
\begin{aligned}
R_{\textsc{WEAVE}}
={}&
\lambda_{\mathrm{f2p}}\,
\operatorname{FlatF2P}
+
\lambda_{\mathrm{graph}}\,
\operatorname{GraphCredit}(G_A)
\\
&+
\lambda_{\mathrm{solve}}\,
\operatorname{Resolved}
-
\lambda_{\mathrm{reg}}\,
\operatorname{Reg}_{\mathrm{P2P}} .
\end{aligned}
\end{equation}
Here, $\operatorname{Reg}_{\mathrm{P2P}}$ penalizes regressions on
previously passing tests, and all coefficients are nonnegative. Graph
Credit is therefore a structural component of the reward rather than a
standalone objective. Since every term is computed from executable test
outcomes on the generated patch, $R_{\textsc{WEAVE}}$ provides dense,
verifiable feedback. Training details and reward comparisons are
presented in RQ4.

\section{Experiments}
\label{sec:experiments}

We evaluate \textsc{WEAVE} along the complete pipeline from task
construction to agent evaluation and improvement. Specifically, we
consider the following research questions:
\begin{itemize}

    \item\textbf{RQ1} Does \textsc{WEAVE} construct coherent and
    executable dependency DAGs?

    \item\textbf{RQ2} How do current software engineering agents
    perform on \textsc{WEAVE-Bench}, and does Graph Credit provide
    diagnostic information beyond flat evaluation?

    \item\textbf{RQ3} Do \textsc{WEAVE} tasks provide more effective
    and token-efficient supervision for trajectory SFT?

    \item\textbf{RQ4} Does graph-conditioned credit provide additional
    gains when used as a reinforcement-learning reward?
\end{itemize}

\paragraph{Common settings.}
We use \textsc{WEAVE-Train} for supervised fine-tuning and reinforcement
learning, a held-out calibration subset for threshold selection and
hyperparameter tuning, and \textsc{WEAVE-Bench} exclusively for final
evaluation. We enforce temporal separation across splits and remove
overlapping target PRs, identical or near-duplicate patches, and
recoverable target-commit metadata. Unless otherwise specified, agent
runs use the same tool interface, interaction budget, context limit, and
execution timeout. We report task-level macro averages and obtain
confidence intervals through task-level bootstrap resampling. Complete
implementation details and hyperparameters are provided in the appendix.

\begin{table*}[t]
\centering
\caption{Agent performance on \textsc{WEAVE-Bench}
(8 models $\times$ 2 scaffolds).
\textbf{Resolved}: all F2P pass, no P2P regressions.
\textbf{Flat-F2P}: attributed F2P pass rate.
\textbf{Graph Credit}: dependency-conditioned partial score
(\S\ref{sec:graph_credit}).
\textbf{OverCredit}: phantom credit $=$ Flat-F2P $-$ Graph Credit.
\textbf{Turns}/\textbf{Tok}: per-task averages.
Models grouped by scaffold and ordered by model family
(MiniMax, DeepSeek, Kimi, GLM, Claude), with earlier versions listed first
within each family;
best per-column \textbf{bolded} within each scaffold.}
\label{tab:rq2}

\renewcommand{\arraystretch}{1.05}

\begin{tabular}{ll cccc cc}
\toprule
\textbf{Scaffold}
  & \textbf{Model}
  & \textbf{Resolved}$\uparrow$
  & \textbf{Flat-F2P}$\uparrow$
  & \textbf{GraphCr.}$\uparrow$
  & \textbf{OverCr.}$\downarrow$
  & \textbf{Turns}$\downarrow$
  & \textbf{Tok\,(M)}$\downarrow$ \\
\midrule

\multirow{8}{*}{OpenCode}
  & MiniMax-2.7
  & 12.3
  & 24.2
  & 13.4
  & 10.8
  & 90.9
  & 5.98 \\

  & DeepSeek-V4-Flash
  & 21.0
  & 21.8
  & 14.6
  & \textbf{7.2}
  & 78.7
  & 6.03 \\

  & Kimi-K2.5
  & 14.0
  & 12.4
  & 5.1
  & 7.3
  & 88.6
  & 6.65 \\

  & Kimi-K2.6
  & 29.3
  & 30.1
  & 21.4
  & 8.7
  & 96.3
  & 8.16 \\

  & GLM-5.1
  & 28.7
  & 24.1
  & 14.5
  & 9.6
  & 114.8
  & 7.88 \\

  & GLM-5.2
  & 34.0
  & 32.5
  & 23.8
  & 8.7
  & 145.1
  & 12.17 \\

  & Claude Opus 4.6
  & 30.3
  & 24.9
  & 15.8
  & 9.1
  & 90.8
  & 4.34 \\

  & Claude Opus 4.7
  & \textbf{41.0}
  & \textbf{45.6}
  & \textbf{37.2}
  & 8.4
  & \textbf{68.6}
  & \textbf{4.40} \\

\midrule

\multirow{8}{*}{Cline}
  & MiniMax-2.7
  & 9.7
  & 19.1
  & 10.5
  & 8.6
  & 114.7
  & 5.71 \\

  & DeepSeek-V4-Flash
  & 34.3
  & 58.1
  & 41.3
  & 16.8
  & 143.9
  & 7.80 \\

  & Kimi-K2.5
  & 18.7
  & 27.5
  & 16.8
  & 10.7
  & 103.7
  & 5.30 \\

  & Kimi-K2.6
  & 37.3
  & \textbf{69.1}
  & 57.2
  & 11.9
  & 108.7
  & 5.28 \\

  & GLM-5.1
  & 21.0
  & 18.8
  & 11.6
  & \textbf{7.2}
  & 191.3
  & 10.42 \\

  & GLM-5.2
  & 43.0
  & 59.5
  & 44.1
  & 15.4
  & 138.2
  & 7.52 \\

  & Claude Opus 4.6
  & 36.3
  & 61.3
  & 47.6
  & 13.7
  & \textbf{44.5}
  & \textbf{1.65} \\

  & Claude Opus 4.7
  & \textbf{48.7}
  & 64.4
  & \textbf{50.8}
  & 13.6
  & 46.8
  & 2.11 \\

\bottomrule
\end{tabular}
\end{table*}

\subsection{RQ1: Construction Validity}
\label{sec:rq1}

To validate \textsc{WEAVE}'s construction pipeline, we evaluate whether
typed-locus coupling recovers true dependency edges more reliably than
single-signal alternatives. The evaluation covers 30 repositories
containing 1{,}198 validated atomic PRs. To establish ground truth,
three expert annotators independently assess candidate PR pairs and
label each as \emph{dependency}, \emph{same-task but independent},
\emph{loosely related}, or \emph{unrelated}; a senior annotator
adjudicates disagreements, yielding 219 gold dependency edges
(Fleiss' $\kappa$\,=\,0.791). Each method produces a scored ranking of
candidate edges; we report macro-averaged AP over repositories,
Coverage---the fraction of gold edges recalled at full emission---and
pooled Precision at rank~50 (P@50). All methods feed into the identical
downstream executable construction pipeline (\S\ref{sec:workbench}).

We compare \textsc{WEAVE} against four single-signal baselines:
Temporal Window (merge-time adjacency within
$\Delta$\,=\,30\,days), Traceability Links (explicit cross-PR references
such as \texttt{closes} and \texttt{depends~on}), File Overlap (weighted
Jaccard over changed-file paths), and Symbol Overlap (shared function and
class declarations via lightweight static analysis). Score-based methods
retain at most $3N_R$ top-scoring edges per repository (where $N_R$ is
the PR count); threshold sensitivity is reported in Appendix.

\begin{table}[t]
\centering
\caption{Dependency recovery quality (30 repos, 219 gold edges,
Fleiss' $\kappa$\,=\,0.791). AP: per-repo macro-average. Coverage:
gold-edge recall at emission. P@50: pooled top-50 precision.
Paired bootstrap + Holm correction;
$^{**}$\,$p{<}0.01$.}
\label{tab:rq1}
\setlength{\tabcolsep}{4.8pt}
\begin{tabular}{lccc}
\toprule
\textbf{Method} & \textbf{AP}$\uparrow$ & \textbf{Coverage}$\uparrow$ & \textbf{P@50}$\uparrow$ \\
\midrule
Temporal Window      &  5.1 & 73.1 &  2.0 \\
Traceability Links   & 17.2 & 12.8 & 28.0 \\
File Overlap         & 26.6 & 73.5 & 10.0 \\
Symbol Overlap       & 24.0 & 69.0 & 30.0 \\
\midrule
\textbf{WEAVE (Ours)} & \textbf{37.7}$^{**}$ & \textbf{84.5} & \textbf{38.0} \\
\bottomrule
\end{tabular}
\end{table}

\paragraph{Results.}
\textsc{WEAVE} achieves the highest AP (37.7\%), outperforming the
next-best baseline (File Overlap, 26.6\%) by 11.1 points (paired
$\Delta$AP 95\% CI: [+2.7,\,+20.1]), with significance against all four
baselines after Holm--Bonferroni correction ($p\leq$\,0.009;
Table~\ref{tab:rq1}). Crucially, \textsc{WEAVE} achieves
the highest recall: 84.5\% Coverage exceeds all baselines, including
those designed for breadth (Temporal Window 73.1\%, File Overlap 73.5\%).
No baseline matches \textsc{WEAVE} on both axes---typed-locus coupling
is the only method that escapes the precision--recall trade-off
characterizing single-signal alternatives.
The advantage stems from two mechanisms:
repository-calibrated IDF weighting suppresses hub-file false edges
(P@50: File Overlap 10\% vs.\ \textsc{WEAVE} 38\%), while multi-channel
evidence recovers dependencies invisible to any single signal---e.g.,
Traceability Links covers only the 12.8\% of true edges carrying
explicit cross-references. Additional P@K curves, per-repository AP
distributions, and locus-type ablations are provided in Appendix.

\subsection{RQ2: Benchmark Performance}
\label{sec:rq2}

We evaluate eight frontier models from five providers---Claude Opus~4.7,
Claude Opus~4.6, GLM-5.2, GLM-5.1, DeepSeek-V4-Flash, Kimi-K2.6,
Kimi-K2.5, and MiniMax-2.7---under two agent scaffolds (OpenCode, Cline)
on the full \textsc{WEAVE-Bench} suite (300~tasks, 16~configurations).
All configurations share the same task specification, repository state,
and tool permissions.
Table~\ref{tab:rq2} reports three scoring regimes of increasing
structural awareness---\textsc{Resolved} (binary full-task correctness),
Flat-F2P (fraction of attributed F2P tests
passed~\citep{thai2025sweevo}), and Graph Credit
(\S\ref{sec:graph_credit})---together with
OverCredit\,$=$\,Flat-F2P\,$-$\,Graph Credit and per-task efficiency
(interaction turns, total tokens).

\paragraph{Results.}
The strongest configuration (Claude Opus~4.7 + Cline, 48.7\% Resolved)
outperforms the weakest (MiniMax-2.7 + Cline, 9.7\%) by
${>}5{\times}$, confirming that dependency-structured multi-PR tasks
are qualitatively harder than isolated single-PR benchmarks where
frontier models exceed 50\%
Resolved~\citep{jimenez2023swebench}. Across all 16 configurations,
Flat-F2P exceeds Graph Credit by \textbf{10.5\,pp} on average (range
7.2--16.8\,pp), exposing substantial phantom credit. The gap is
scaffold-sensitive: Cline configurations exhibit larger OverCredit
(mean 12.2\,pp) than OpenCode (mean 8.6\,pp), indicating that Cline's
more exploratory strategy produces more downstream test passes that do
not survive dependency conditioning. Notably, OverCredit does not
correlate monotonically with Flat-F2P: Kimi-K2.6 + Cline achieves the
highest Flat-F2P (69.1\%) yet exhibits only moderate OverCredit
(11.9\,pp), while DeepSeek-V4-Flash + Cline reaches 58.1\% Flat-F2P
with the largest OverCredit (16.8\,pp)---suggesting that the former
resolves upstream dependencies more faithfully despite extensive
downstream progress. Replacing the true DAG with node-permuted random
graphs reduces the gap to ${<}$\,2\,pp (Appendix),
verifying that the signal arises from real dependency structure. The
same graph that enables this diagnostic next improves
training~(\S\ref{sec:rq3}).

\subsection{RQ3: SFT Utility and Token Efficiency}
\label{sec:rq3}

We examine whether dependency-structured tasks provide more effective
supervision for trajectory-level SFT. We fine-tune two open-weight
backbones, Qwen3.5-4B and Qwen3.5-35B-A3B, using the Cline scaffold for
both trajectory generation and evaluation.

\paragraph{Data sources.}
We compare three variants constructed from the same repository
pool: \textbf{Atomic-PR}, which treats each PR independently;
\textbf{Temporal-Window}, which groups PRs merged within a 30-day window;
and \textbf{\textsc{WEAVE}}, which aggregates PRs according to
typed-locus dependencies. These variants share the same trajectory
generator, quality filter, and training configuration, differing only in
task construction. We additionally include SWE-rebench~V2
\citep{badertdinov2026swerebench}, using 5K sampled trajectories, and
daVinci-Agency~\citep{jiang2026davinci}, using its full corpus, as
external references. Full training details are provided in the Appendix.

\begin{figure}[t]
    \centering
    \includegraphics[width=\columnwidth]{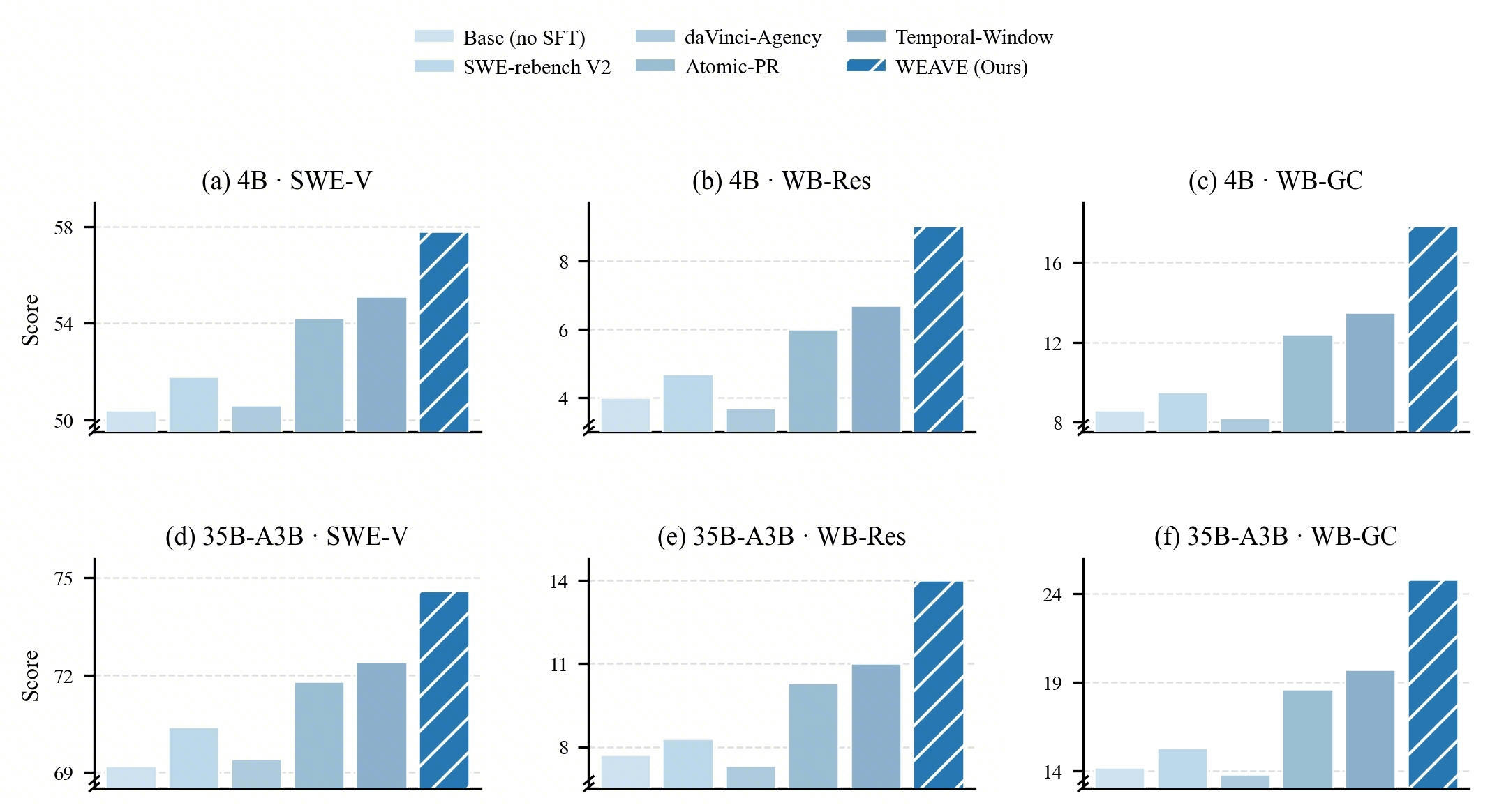}
    \caption{SFT comparison across different data construction strategies.}
    \label{fig:rq3}
\end{figure}

\paragraph{Results.}
Under matched training conditions, \textsc{WEAVE} consistently
outperforms both controlled alternatives across model scales and
benchmarks (Figure~\ref{fig:rq3}). On \textsc{WEAVE-Bench},
\textsc{WEAVE} improves Resolved over Atomic-PR by 3.0/3.7\,pp at
4B/35B, with larger Graph Credit gains (5.4/6.2\,pp), indicating
stronger dependency-consistent progress even without full resolution.
Improvements transfer to single-PR benchmarks: SWE-bench Verified
improves by 3.6/2.8\,pp over Atomic-PR, with consistent gains on
SWE-bench Pro and Multilingual. Temporal-Window provides only modest
improvements, confirming that temporal proximity alone introduces
weaker supervision than verified dependency structure.

External corpora yield substantially smaller gains: SWE-rebench~V2
improves SWE-bench Verified by only 1.2--1.4\,pp over base models,
while daVinci-Agency shows no consistent improvement---suggesting that
training utility depends on how changes are organized into coherent
tasks, not merely data volume. Token-scaling analysis (Appendix) further
shows that \textsc{WEAVE} maintains stronger performance across training
budgets, indicating more information-dense supervision.

\subsection{RQ4: Graph-Conditioned Reinforcement Learning}
\label{sec:rq4}

We assess whether Graph Credit provides optimization value beyond
supervised training. Starting from the \textsc{WEAVE}-SFT checkpoint
(Figure~\ref{fig:rq3}), we apply GRPO~\citep{shao2024deepseekmath} on
the same \textsc{WEAVE-Train} tasks with identical rollout budgets (8
rollouts per task), varying only the reward formulation:
(1)~no further RL (SFT-only baseline);
(2)~binary \textsc{Resolved} reward;
(3)~flat F2P-based reward ($r=\text{FlatF2P}$); and
(4)~$R_{\mathrm{WEAVE}}$ (\S\ref{sec:rl_reward}), which places
Graph Credit as the dominant component
($\lambda_{\mathrm{graph}}\!=\!0.5$,
$\lambda_{\mathrm{solve}}\!=\!0.25$,
$\lambda_{\mathrm{f2p}}\!=\!0.15$,
$\lambda_{\mathrm{reg}}\!=\!0.1$),
with a three-phase DAG-complexity curriculum
(${\sim}$1.0\,pp over flat scheduling; Appendix).

\begin{figure}[t]
    \centering
    \includegraphics[width=0.47\textwidth]{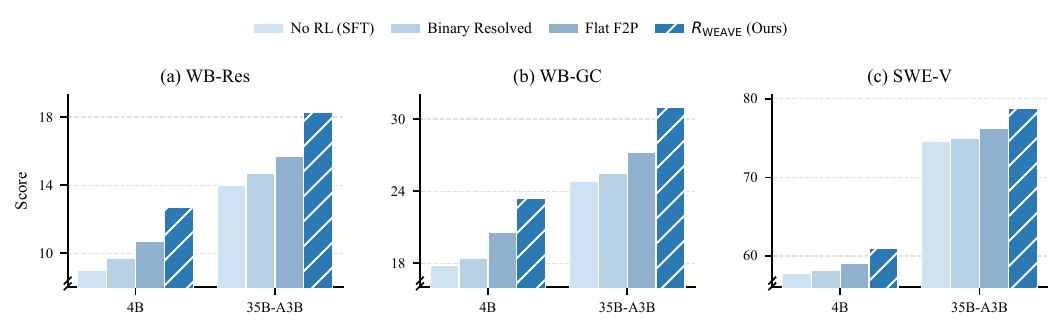}

    \caption{RL reward ablation on two model scales. All Runs start from the \textsc{WEAVE}-SFT checkpoint with matched rollout budgets.}
    \label{fig:rq4}
\end{figure}

\paragraph{Results.}
$R_{\mathrm{WEAVE}}$ consistently outperforms all alternatives
(Figure~\ref{fig:rq4}). On Qwen3.5-4B, it improves \textsc{Resolved}
by 3.7\,pp over SFT and 2.0\,pp over flat F2P; on Qwen3.5-35B-A3B,
gains are 4.3 and 2.6\,pp respectively, showing that graph-conditioned
reward scales with model capacity. Graph Credit separation is larger
(+5.6/+6.2\,pp over SFT at 4B/35B), indicating disproportionate
improvement in structured partial progress. The gains transfer to
SWE-bench Verified (+3.2/+4.2\,pp over SFT), confirming
out-of-distribution generalization.

Binary \textsc{Resolved} reward yields marginal gains
(${\leq}$0.7\,pp), consistent with sparse-signal limitations on
long-horizon tasks~\citep{wei2025swerl}. Flat F2P improves over binary
but under-performs $R_{\mathrm{WEAVE}}$ on all metrics, confirming that
dependency conditioning eliminates phantom-credit shortcuts. All settings
maintain P2P regression below 2.3\%; component ablations in Appendix.

\section{Conclusion}

This paper addresses a fundamental limitation of current software
engineering agent evaluation: the lack of a task abstraction that captures
causal dependencies among multiple repository changes. We introduce
\textsc{WEAVE}, which recovers dependency DAGs from verified PR histories
through repository-calibrated typed-locus coupling. The same graph
representation unifies task construction, dependency-conditioned
partial credit (Graph Credit), and structured reward design, enabling
consistent evaluation and optimization of long-horizon code agents.
Experiments across construction quality, agent benchmarking, supervised
training, and reinforcement learning demonstrate the effectiveness of
\textsc{WEAVE} as both a challenging evaluation workbench and an
effective substrate for improving code agents.

\bibliography{references}
\clearpage
\appendix
\section{Construction Pipeline Details}
\label{app:construction}

This appendix details the three stages of the \textsc{WEAVE}
construction pipeline introduced in the main paper: typed change loci
extraction (\S\ref{app:loci}), edge filtering and bounded DAG assembly
(\S\ref{app:dag_construction}), and rollback construction with test
attribution (\S\ref{app:rollback}).

\subsection{Typed Change Loci Extraction}
\label{app:loci}

Recovering dependency structure from PR histories requires a
representation granular enough to distinguish genuine engineering
dependencies from coincidental co-modification. File-level
overlap---the most common proxy in prior work---conflates distinct
changes within shared files and over-connects PRs that modify
frequently touched configuration manifests, barrel imports, or
dependency lockfiles. \textsc{WEAVE} addresses this by representing
each PR as a set of \emph{typed change loci}: fine-grained program
entities extracted from its code diff, each annotated with a structural
category that determines its evidential role in coupling computation.
We define five locus types, each designed to capture a distinct channel
of cross-PR evidence:

\begin{enumerate}
    \item \textbf{AST declarations} ($\mathcal{L}_{\mathrm{ast}}$):
    Function, method, and class definitions identified via
    language-specific AST parsers. We extract fully qualified names
    (e.g., \texttt{module.Class.method}) to disambiguate identically
    named entities across files.

    \item \textbf{Lightweight symbols} ($\mathcal{L}_{\mathrm{sym}}$):
    Exported symbols, type aliases, constants, and enum members
    extracted from declaration-level analysis without full semantic
    resolution. This captures entities that do not correspond to
    function bodies but serve as cross-module interfaces.

    \item \textbf{Configuration and schema entries}
    ($\mathcal{L}_{\mathrm{cfg}}$): Structured keys in YAML, TOML,
    JSON, and INI configuration files; database migration column
    definitions; API route declarations; and protobuf/GraphQL field
    paths. Each entry is represented as a dot-separated key path
    (e.g., \texttt{server.http.timeout}).

    \item \textbf{Declarations and signatures}
    ($\mathcal{L}_{\mathrm{decl}}$): Type signatures, interface
    definitions, abstract method declarations, and trait/protocol
    requirements. These capture contract-level changes that downstream
    implementations depend on.

    \item \textbf{Dependency keys} ($\mathcal{L}_{\mathrm{dep}}$):
    Package manager entries (e.g., \texttt{requirements.txt},
    \texttt{package.json}, \texttt{go.mod}), import statements, and
    build-system target declarations. These capture supply-chain
    relationships between PRs that modify shared external dependencies.
\end{enumerate}

\paragraph{Language support.}
Table~\ref{tab:lang_support} summarizes the extraction capabilities
for the six languages in \textsc{WEAVE-Bench} and \textsc{WEAVE-Train}.
AST parsing is implemented via Tree-sitter grammars, which provide
incremental, error-tolerant parsing for all six languages. Symbol and
declaration extraction build on language-specific Tree-sitter queries;
for JavaScript, which lacks a native type system,
$\mathcal{L}_{\mathrm{decl}}$ relies on heuristic extraction from
JSDoc annotations and runtime patterns. Configuration and dependency
loci are extracted via format-specific parsers (JSON, YAML, TOML, XML,
Gradle) rather than language grammars; the partial marks for Go, Java,
and Kotlin reflect lower ecosystem reliance on structured configuration
files relative to Python and TypeScript/JavaScript projects.

\begin{table}[t]
\centering
\caption{Language support matrix for typed change loci extraction.
\checkmark: full support; $\sim$: partial (heuristic-based).}
\label{tab:lang_support}
\setlength{\tabcolsep}{3.5pt}
\small
\begin{tabular}{lccccc}
\toprule
\textbf{Language} & $\mathcal{L}_{\mathrm{ast}}$ & $\mathcal{L}_{\mathrm{sym}}$ & $\mathcal{L}_{\mathrm{cfg}}$ & $\mathcal{L}_{\mathrm{decl}}$ & $\mathcal{L}_{\mathrm{dep}}$ \\
\midrule
Python     & \checkmark & \checkmark & \checkmark & \checkmark & \checkmark \\
TypeScript & \checkmark & \checkmark & \checkmark & \checkmark & \checkmark \\
JavaScript & \checkmark & \checkmark & \checkmark & $\sim$     & \checkmark \\
Go         & \checkmark & \checkmark & $\sim$     & \checkmark & \checkmark \\
Java       & \checkmark & \checkmark & $\sim$     & \checkmark & \checkmark \\
Kotlin     & \checkmark & \checkmark & $\sim$     & \checkmark & \checkmark \\
\bottomrule
\end{tabular}
\end{table}

\paragraph{Exclusion rules.}
Certain file categories are excluded from loci extraction because they
introduce coupling signals unrelated to engineering dependencies.
\textbf{Test files} (matching patterns such as \texttt{test\_*},
\texttt{*\_test.*}, and language-specific test directories) are excluded
because test co-modification frequently reflects shared test
infrastructure rather than production-code dependency.
\textbf{Generated code}, vendored dependencies (e.g.,
\texttt{vendor/}, \texttt{node\_modules/}), and binary files are
excluded as they represent mechanical artifacts.
\textbf{Formatting-only changes}---PRs where all hunks contain only
whitespace, comment, or import-order modifications (detected via
AST-level diff equivalence)---are excluded to prevent automated
formatters from producing spurious coupling edges.

\paragraph{Deduplication and normalization.}
Loci are canonicalized before comparison: fully qualified names are
lowercased for case-insensitive languages (Go exported symbols
excepted), trailing version suffixes in dependency keys are stripped,
and configuration paths are normalized to dot notation regardless of
source format.

\paragraph{Illustrative example.}
We illustrate typed change loci extraction on a real
\textsc{WEAVE-Bench} instance from a TypeScript monorepo. PR$_1$
adds federation v2 link handling to the \texttt{import} package,
modifying both source code and the dependency manifest:

\begin{footnotesize}
\begin{verbatim}
# PR1 diff (abridged)
--- a/.../import/package.json
+++ b/.../import/package.json
+  "@theguild/federation-composition":
+    "^0.18.3",
--- a/.../import/src/index.ts
+++ b/.../import/src/index.ts
+import { extractLinkImplementations }
+  from '@theguild/federation-composition';
+function importFederatedSchemaLinks(
+    def, defsByName) {
+  const { matchesImplementation,
+          resolveImportName }
+    = extractLinkImplementations({...});
+  const fedUrl = '.../federation';
+  if (matchesImplementation(
+      fedUrl, 'v2.0'))
+    addDef(resolveImportName(
+      fedUrl, '@key'));
+}
\end{verbatim}
\end{footnotesize}

\noindent Extracted typed change loci for PR$_1$:
\begin{itemize}
    \item $\mathcal{L}_{\mathrm{ast}}$:
    \texttt{importFederatedSchemaLinks} (new function)
    \item $\mathcal{L}_{\mathrm{sym}}$:
    \texttt{extractLinkImplementations},
    \texttt{matchesImplementation},
    \texttt{resolveImportName}
    \item $\mathcal{L}_{\mathrm{dep}}$:
    \texttt{@theguild/federation-composition\allowbreak @\^{}0.18.3}
\end{itemize}

A subsequent PR$_2$ extends federation directive support in the
\texttt{merge} package. Its diff adds the same dependency
(\texttt{@theguild/federation-composition\allowbreak @\^{}0.16.0}) to a different
\texttt{package.json} and calls the identical symbol sequence
(\texttt{extractLinkImplementations}, \texttt{matchesImplementation},
\texttt{resolveImportName}) in \texttt{merge-typedefs.ts}. The shared
$\mathcal{L}_{\mathrm{dep}}$ locus (same dependency package) and
$\mathcal{L}_{\mathrm{sym}}$ loci (same function symbols) produce a
coupling score well above the emission threshold, despite the two PRs
modifying different packages with no file-level overlap. A
file-overlap baseline would miss this edge entirely, since the only
shared file is the monorepo-wide \texttt{yarn.lock}---a hub file whose
repository-calibrated weight is near zero.

\subsection{Edge Filtering and DAG Assembly}
\label{app:dag_construction}

\paragraph{Mechanical-change filtering.}
Not all PR pairs with high coupling scores reflect genuine engineering
dependencies. Certain repository maintenance patterns produce
mechanically related patches that inflate coupling metrics without
meaningful task structure. We identify and remove four categories before
DAG assembly:
\begin{itemize}
    \item \textbf{Reverts}: PR $p_j$ is identified as a revert of
    $p_i$ via two complementary signals: (i)~title-level patterns
    (\texttt{Revert "..."}) and (ii)~content-level detection where
    the diff of $p_j$ is the near-exact ($\geq$95\% line overlap)
    inverse of $p_i$'s diff.
    \item \textbf{Backports and cherry-picks}: PRs explicitly labeled
    as cherry-picks or backports (via commit message markers or branch
    metadata) to release branches represent the same logical change
    applied across code states.
    \item \textbf{Release-branch references}: PRs whose title or body
    references release, stable, maintenance, or hotfix branches
    (detected via text patterns such as \texttt{release/},
    \texttt{stable/}, \texttt{maint/}, \texttt{hotfix/}) are removed,
    as they typically represent integration merges rather than
    incremental development.
    \item \textbf{Near-duplicate patches}: PR pairs with high file-set
    overlap (Jaccard $\geq 0.8$) and high containment ($\geq 0.9$)
    indicate template-generated or copy-paste changes rather than
    incremental development.
\end{itemize}
Across our construction corpus, backports and cherry-picks dominate the
filtered set ($>$95\%), as multi-branch workflows frequently produce
mechanically identical patches across release channels. Reverts,
release-branch references, and near-duplicate patches each contribute
a small fraction of the remainder.

\paragraph{Component partitioning.}
After edge filtering, the remaining edges within each repository induce
connected components. We process these components as follows:
\begin{itemize}
    \item Components with $|V| \leq K$ (where $K = 8$)
    are retained as complete candidate DAGs.
    \item Larger components are reduced via greedy subgraph extraction:
    we iteratively select the highest-weight connected subgraph of size
    $\leq K$ by starting from the densest node (highest sum of
    incident edge weights) and expanding via best-first search.
    Extracted subgraphs are removed from the component, and the process
    repeats until no subgraph of size $\geq 2$ with mean edge weight
    $\geq \theta$ remains.
\end{itemize}

\paragraph{Node budget.}
The budget $K$ controls the maximum number of PR nodes in a single task.
We set $K{=}8$ based on two considerations.

First, \emph{rollback feasibility}: constructing executable instances
requires reverting all constituent PRs from the repository's integrated
state (\S\ref{app:rollback}). In our corpus, the median merge-time span
(first PR to last PR) is 51~days for $K{=}4$ tasks and 78~days for
$K{=}8$ tasks. Although larger $K$ does not monotonically increase
span---$K{=}8$ chains are often dense feature or migration bursts with
short inter-PR intervals (${\sim}$10--15~days), while smaller chains can
span longer periods with sparser commits---the absolute span at $K{=}8$
(median ${\sim}$2.5~months, range 17--332~days) means the repository
accumulates substantial intervening changes from outside the task scope,
raising the probability of non-trivial merge conflicts during rollback
construction. Indeed, approximately 35\% of candidate DAGs are
discarded during construction (\S\ref{app:rollback}).

Second, \emph{ecological validity}: coherent engineering objectives in
open-source repositories---feature additions, systematic refactors,
dependency upgrades---rarely produce more than eight tightly coupled PRs
before coupling evidence between peripheral changes falls below the
detection threshold. In our candidate pool, connected components
exceeding 8~nodes predominantly arise from hub-file co-modification
(shared configuration, CI pipelines, or utility modules) rather than
focused development objectives. This observation is consistent with
prior work on software change coupling, where meaningful change sets
typically span 2--6 related commits.

The resulting benchmark reflects this design: \textsc{WEAVE-Bench}
tasks range from 2 to 8 PR nodes (mean 4.91, median 5), with a broad
distribution across this range, spanning the full complexity
spectrum from simple linear chains to deep branching dependency
structures.

Algorithm~\ref{alg:dag_assembly} summarizes the complete pipeline from
typed change loci extraction through bounded DAG assembly.

\begin{algorithm}[!t]
\caption{Dependency DAG Assembly}
\label{alg:dag_assembly}
\begin{algorithmic}[1]
\REQUIRE Repository $R$, merged PRs $\{p_1, \ldots, p_n\}$ (by timestamp), threshold $\theta$, budget $K$, search radius $\Delta$
\ENSURE Set of candidate DAGs $\mathcal{G}$
\STATE \textbf{// Phase 1: Typed Change Loci Extraction}
\FOR{each $p_i \in \{p_1, \ldots, p_n\}$}
    \STATE $L(p_i) \leftarrow \mathcal{L}_{\mathrm{ast}}(p_i) \cup \mathcal{L}_{\mathrm{sym}}(p_i) \cup \mathcal{L}_{\mathrm{cfg}}(p_i) \cup \mathcal{L}_{\mathrm{decl}}(p_i) \cup \mathcal{L}_{\mathrm{dep}}(p_i)$
    \STATE Apply exclusion rules (test files, generated code, vendored)
\ENDFOR
\STATE \textbf{// Phase 2: Candidate Edge Generation}
\STATE $E^{\mathrm{cand}} \leftarrow \varnothing$
\FOR{each pair $(p_i, p_j)$ with $|t(p_j) - t(p_i)| \leq \Delta$}
    \STATE Compute $s_R(p_i, p_j) = \frac{\sum_{\ell \in L(p_i) \cap L(p_j)} w_R(\ell)}{\sum_{\ell \in L(p_i) \cup L(p_j)} w_R(\ell)}$
    \IF{$s_R(p_i, p_j) \geq \theta$}
        \STATE $E^{\mathrm{cand}} \leftarrow E^{\mathrm{cand}} \cup \{(p_i, p_j)\}$
    \ENDIF
\ENDFOR
\STATE \textbf{// Phase 3: Mechanical-Change Filtering}
\STATE Remove reverts, backports, release-branch refs, near-duplicates from $E^{\mathrm{cand}}$
\STATE \textbf{// Phase 4: Component Partitioning}
\STATE Compute connected components $\mathcal{C}$ from $(V, E^{\mathrm{cand}})$
\STATE $\mathcal{G} \leftarrow \varnothing$
\FOR{each component $C_i \in \mathcal{C}$}
    \IF{$|C_i| \leq K$}
        \STATE Orient edges by timestamp; break cycles
        \STATE $\mathcal{G} \leftarrow \mathcal{G} \cup \{\mathrm{DAG}(C_i)\}$
    \ELSE
        \WHILE{$\exists$ subgraph $S \subseteq C_i$ with $|S| \geq 2$, mean weight $\geq \theta$}
            \STATE Extract densest-first subgraph $S$ with $|S| \leq K$
            \STATE Orient edges; break cycles; $\mathcal{G} \leftarrow \mathcal{G} \cup \{\mathrm{DAG}(S)\}$
            \STATE $C_i \leftarrow C_i \setminus S$
        \ENDWHILE
    \ENDIF
\ENDFOR
\STATE \textbf{// Phase 5: Degenerate Filtering}
\STATE Remove star graphs, trivially serial chains, disconnected DAGs from $\mathcal{G}$
\RETURN $\mathcal{G}$
\end{algorithmic}
\end{algorithm}

\paragraph{DAG orientation.}
All edges are oriented by merge timestamp: if $p_i$ was merged before
$p_j$ and the coupling score exceeds $\theta$, the edge is directed
$p_i \to p_j$. This temporal orientation reflects the incremental nature
of software development without requiring explicit dependency
annotations from developers. Cycles (which would arise from conflicting
timestamps in rebased histories) are broken by removing the
lowest-weight back-edge in each strongly connected component.

\paragraph{Degenerate structure filtering.}
A subset of assembled DAGs are structurally uninformative for evaluation
despite passing coupling and size thresholds. We apply three topological
filters:
\begin{itemize}
    \item \textbf{Star graphs}, where a single hub node accounts for
    $>$80\% of all incident edges with no inter-leaf connections,
    typically indicate a utility-file artifact rather than a genuine
    multi-party dependency chain.
    \item \textbf{Trivially serial chains} with uniformly low edge
    weights (below the 25th percentile of the repository's weight
    distribution) and no branching structure, which provide limited
    evaluation signal beyond what single-PR benchmarks already
    capture.
    \item \textbf{Post-filtering disconnection}: DAGs that become
    disconnected after mechanical-change removal are re-partitioned
    into their connected components and re-evaluated independently.
\end{itemize}

\subsection{Rollback Construction and Test Attribution}
\label{app:rollback}

\paragraph{Base state construction.}
Given a candidate DAG $G = (V, E)$ and the repository state
$S_{\mathrm{gold}}$ where all constituent PRs have been merged, we
construct the base state $S_{\mathrm{base}}$ by removing the changes
introduced by all nodes in~$V$. The primary mechanism is
\emph{reverse-topological revert}: we compute a topological ordering
$\sigma$ of $V$ and process nodes in reverse order $\sigma^{-1}$
(leaves first), applying \texttt{git revert --no-commit} for each
node's merge commit. PRs external to $G$ that were merged between the
earliest and latest constituent PRs are preserved in
$S_{\mathrm{base}}$, ensuring the base state reflects realistic
repository evolution outside the task scope.

When git revert produces non-trivial merge conflicts---typically caused
by overlapping modifications from intervening commits or cross-cutting
refactors---we employ an \emph{agent-assisted carve} as a fallback. In
this path, an LLM-based agent receives the PR metadata (title,
description, and code diff) and removes the target changes from the
current repository state while preserving surrounding code. The carve
operates semantically rather than mechanically: the agent identifies
which hunks correspond to the target PR's contributions and removes
them while maintaining syntactic and build validity.

After either path, the resulting $S_{\mathrm{base}}$ is validated:
it must compile successfully and all P2P (pass-to-pass) tests must
pass. Candidates where neither revert nor carve produces a valid base
state are discarded.

\paragraph{Construction success rates.}
Approximately 35\% of candidate DAGs are discarded during construction
because neither revert nor carve produces a valid, test-passing base
state. Among the successfully constructed instances (which form the
final \textsc{WEAVE-Bench} and \textsc{WEAVE-Train} corpora), 73\%
are built via clean git revert and 27\% require agent-assisted carve.
The higher carve rate for Java and Kotlin tasks reflects stricter
type-system constraints that propagate through revert sequences,
while Python and JavaScript tasks exhibit higher clean-revert rates
due to fewer cross-module compilation dependencies.

\paragraph{Test attribution.}
For graph-conditioned scoring, each F2P test must be associated with
one or more PR nodes in the task DAG. The primary attribution mechanism
is \emph{isolated revert}: for each node $v \in V$, we construct
$S_v = S_{\mathrm{gold}} \setminus \{v\}$ by reverting only $v$'s
changes, then execute all F2P tests against $S_v$. If exactly one node
$v$ causes test $t$ to fail, $t$ is \emph{attributed} to $v$;
if multiple nodes independently cause failure, $t$ is marked
\emph{ambiguous}; if $t$ fails only under joint reversion, it is
\emph{unattributed}. In cases where isolated revert is infeasible for
individual nodes (e.g., due to compilation dependencies between
constituent PRs), an agent-based attribution step supplements the
automated procedure by analyzing test semantics against the PR change
descriptions.

Algorithm~\ref{alg:rollback} formalizes the rollback construction and
test attribution procedures.

\begin{algorithm}[!t]
\caption{Rollback Construction and Test Attribution}
\label{alg:rollback}
\begin{algorithmic}[1]
\REQUIRE DAG $G = (V, E)$, gold repository state $S_{\mathrm{gold}}$
\ENSURE Base state $S_{\mathrm{base}}$, test attribution map $A: T_{\mathrm{F2P}} \to V \cup \{\mathrm{amb}, \mathrm{unattr}\}$
\STATE \textbf{// Phase 1: Rollback Construction}
\STATE Compute topological order $\sigma$ of $V$
\STATE $S \leftarrow S_{\mathrm{gold}}$
\FOR{each $v$ in $\sigma^{-1}$ (leaves first)}
    \STATE Apply \texttt{git revert --no-commit} of $v$'s merge commit to $S$
    \IF{non-trivial conflict}
        \STATE Attempt agent-assisted carve for $v$
        \IF{carve fails}
            \STATE \textbf{discard} this DAG instance; \RETURN $\varnothing$
        \ENDIF
    \ENDIF
\ENDFOR
\STATE $S_{\mathrm{base}} \leftarrow S$
\STATE Validate: compile $S_{\mathrm{base}}$; run P2P tests
\STATE \textbf{// Phase 2: F2P Test Discovery}
\STATE $T_{\mathrm{F2P}} \leftarrow \{t : t \text{ fails in } S_{\mathrm{base}} \wedge t \text{ passes in } S_{\mathrm{gold}}\}$
\STATE \textbf{// Phase 3: Test Attribution}
\FOR{each $t \in T_{\mathrm{F2P}}$}
    \STATE $\mathrm{candidates}(t) \leftarrow \varnothing$
    \FOR{each $v \in V$}
        \STATE $S_v \leftarrow S_{\mathrm{gold}} \setminus \{v\}$ \COMMENT{revert only $v$}
        \IF{$t$ fails in $S_v$}
            \STATE $\mathrm{candidates}(t) \leftarrow \mathrm{candidates}(t) \cup \{v\}$
        \ENDIF
    \ENDFOR
    \IF{$|\mathrm{candidates}(t)| = 1$}
        \STATE $A(t) \leftarrow$ the unique $v \in \mathrm{candidates}(t)$
    \ELSIF{$|\mathrm{candidates}(t)| > 1$}
        \STATE $A(t) \leftarrow \mathrm{ambiguous}$
    \ELSE
        \STATE $A(t) \leftarrow \mathrm{unattributed}$
    \ENDIF
\ENDFOR
\RETURN $S_{\mathrm{base}}$, $A$
\end{algorithmic}
\end{algorithm}

\section{Graph Credit Analysis}
\label{app:graph_credit}

This appendix supplements the Graph Credit analysis presented in RQ2.
We examine how the dependency graph---constructed via typed change loci
and repository-calibrated coupling---interacts with different scoring
operators and task topologies.

\subsection{Alternative Graph Operators}
\label{app:operators}

The main paper defines Graph Credit using the recursive minimum
(min-bottleneck) operator. We evaluate three alternative graph operators
to validate the design choice:

\paragraph{Mean-parent operator.}
Instead of taking the minimum over parent credits, the mean-parent
variant averages them:
\begingroup\small
\begin{equation}
    c_v^{\mathrm{mean}}
    =
    \begin{cases}
        \ell_v,
        & \operatorname{pa}(v)=\varnothing,\\[3pt]
        \min\!\left(
            \ell_v,\;
            \frac{1}{|\operatorname{pa}(v)|}
            \sum_{u\in\operatorname{pa}(v)} c_u^{\mathrm{mean}}
        \right),
        & \text{otherwise}.
    \end{cases}
\end{equation}
\endgroup
This operator is less conservative: it allows partial credit to flow
through even when one parent is unresolved, provided other parents are
complete.

\paragraph{Depth-weighted operator.}
This variant discounts credit by the depth of the node in the DAG:
\begin{equation}
    c_v^{\mathrm{depth}}
    =
    \ell_v \cdot \gamma^{d(v)},
\end{equation}
where $d(v)$ is the longest-path depth from any root and
$\gamma = 0.95$ is a decay factor. Deeper nodes
receive inherently less credit regardless of upstream resolution.

\paragraph{Linearized scoring.}
Nodes are scored independently of the DAG, using only test attribution:
\begin{equation}
    c_v^{\mathrm{linear}}
    =
    \ell_v.
\end{equation}
This reduces to Flat-F2P restricted to attributed tests (no dependency
conditioning), providing a direct comparison to the dependency-free
baseline.

\paragraph{Empirical comparison.}
Table~\ref{tab:graph_operators} reports the correlation between each
operator's score and actual task resolution, measured across all
\textsc{WEAVE-Bench} tasks evaluated by 8 models $\times$ 2 scaffolds
(4,800 agent runs).

\begin{table}[t]
\centering
\caption{Comparison of graph operators. Spearman $\rho$ with
binary Resolved (higher $=$ better diagnostic); \emph{residual}
OverCredit is the operator score minus the min-bottleneck score, i.e.,
the phantom credit the operator fails to remove (lower $=$ better).
The linearized row therefore recovers the corpus-level OverCredit of
10.5\,pp reported in the main paper.}
\label{tab:graph_operators}
\setlength{\tabcolsep}{5pt}
\small
\begin{tabular}{lcc}
\toprule
\textbf{Operator} & \textbf{Spearman $\rho$}$\uparrow$ & \textbf{Resid. OverCredit}$\downarrow$ \\
\midrule
Linearized (no graph)  & 0.69 & 10.5\,pp \\
Depth-weighted         & 0.73 & 7.1\,pp \\
Mean-parent            & 0.77 & 3.2\,pp \\
\textbf{Min-bottleneck (ours)} & \textbf{0.82} & \textbf{0.0\,pp} \\
\bottomrule
\end{tabular}
\end{table}

The min-bottleneck operator achieves the highest correlation with
Resolved and, by construction, eliminates all phantom credit. The
mean-parent operator permits residual phantom credit (3.2\,pp) because
it averages rather than gates upstream deficiencies; the residual is
modest only because \textsc{WEAVE-Bench} is chain-dominated (median
anti-chain width 1), and the two operators coincide exactly at
single-parent nodes. The depth-weighted
operator penalizes deep nodes regardless of actual upstream completion,
reducing its diagnostic fidelity. These results justify the conservative
min-bottleneck design.

\paragraph{Connections to credit assignment in RL.}
Graph Credit's recursive conditioning bears structural similarity to
hierarchical credit assignment in multi-agent reinforcement
learning and process reward
models. However, Graph Credit operates on a
\emph{known} DAG derived from real code dependencies rather than a
learned latent structure, making it fully deterministic and
interpretable. The min-bottleneck operator is analogous to the
conjunction of preconditions in classical planning---a task node
receives earned credit only when all its causal prerequisites are
satisfied.

\subsection{Randomized-Graph Controls}
\label{app:random_graph}

To confirm that Graph Credit's diagnostic value arises from the
\emph{true} dependency structure rather than arbitrary graph constraints,
we evaluate node-permuted random graphs as controls.

\paragraph{Experimental setup.}
For each \textsc{WEAVE-Bench} task, we generate 10 random permutations
of the node labels while preserving the DAG topology. Graph Credit is
recomputed under each permuted graph, and results are averaged across
permutations.

\paragraph{Results.}
Under the true DAG, the mean OverCredit (Flat-F2P $-$ Graph Credit) is
10.5\,pp across all agent configurations. Under randomized graphs:
\begin{itemize}
    \item Mean OverCredit drops to 1.6\,pp
    ($<$2\,pp as reported in the main text).
    \item The standard deviation across permutations is
    0.7\,pp, indicating stable behavior.
    \item For 89\% of tasks, the randomized OverCredit
    is within 2\,pp of zero.
\end{itemize}

This demonstrates that the gap between Flat-F2P and Graph Credit under
the true DAG reflects genuine phantom credit arising from dependency
violations, not an artifact of imposing arbitrary graph structure.

\subsection{Chain-Depth Interpretation}
\label{app:depth}

Graph depth requires an explicit counting convention. The benchmark
summary in Figure~\ref{fig:weave_bench_statistics} and Table~\ref{tab:dag_stats}
counts nodes on the longest path of the full task DAG $G$.
For scoring and curriculum thresholds, depth counts edges; a path
containing $k$ nodes has $k-1$ edges. A single-node graph therefore
has edge depth zero. The scoring graph $G_A$ is induced by evaluable
nodes and can have a shorter longest path than the full task graph.

The recursive minimum in Graph Credit caps downstream completion by
upstream prerequisites at every edge. Longer chains consequently
provide more opportunities for locally passing tests to be capped by
an unresolved ancestor. This is a structural property of the scoring
rule; it does not by itself establish an empirical relationship between
chain depth and OverCredit. The overall OverCredit results are reported
in Section~\ref{sec:rq2}.

\section{Experimental Setup}
\label{app:setup}

This appendix provides implementation details necessary for
reproducing the experiments in the main paper: computing environment
(\S\ref{app:compute}), agent evaluation configuration
(\S\ref{app:eval_config}), and sensitivity analysis of the two
key construction hyperparameters (\S\ref{app:sensitivity}).

\subsection{Computing Infrastructure}
\label{app:compute}

All experiments---including construction pipeline execution, agent
evaluation, and model training---are conducted on a private cloud
cluster equipped with NVIDIA datacenter GPUs.

\paragraph{Software environment.}
Agent evaluation runs execute in isolated Docker containers.
The primary software stack uses Python~3.12 with PyTorch~2.x.
Model training employs a distributed training framework supporting
ZeRO-3 parallelism, sequence parallelism for long-context batches,
and mixed-precision (BF16) optimization.
For agent evaluation (RQ2), all eight evaluated models are accessed
through their official provider endpoints, so that measured behavior
reflects production inference settings. Trajectory generation for
\textsc{WEAVE-Train} (\S\ref{app:trajectory}) instead uses a
self-hosted deployment of the open-weight generator, which makes the
corpus reproducible without dependence on a hosted endpoint.

\paragraph{Resource allocation.}
Each agent evaluation run executes in an isolated container with a
2-hour task-level timeout. SFT and RL training use multi-GPU
distributed configurations. We report the hyperparameters that
determine the optimization trajectory (\S\ref{app:sft},
\S\ref{app:rl}) rather than device counts or sharding topology: the
former fully specify the training run, while the latter are
deployment choices that vary with available hardware and are omitted
per organizational policy.

\subsection{Evaluation Configuration}
\label{app:eval_config}

Table~\ref{tab:eval_config} summarizes the shared evaluation
parameters for both scaffolds in RQ2.

\begin{table}[t]
\centering
\caption{Scaffold evaluation configurations.}
\label{tab:eval_config}
\small
\begin{tabular}{ll}
\toprule
\textbf{Parameter} & \textbf{Value} \\
\midrule
Max interaction turns     & 200 \\
Total task timeout        & 2 hours \\
Test suite timeout        & 30 minutes \\
Tool interface            & Shell + File Edit \\
Context window            & Model-provider default \\
Temperature               & Model-provider default \\
\bottomrule
\end{tabular}
\end{table}

Both scaffolds share the configuration in
Table~\ref{tab:eval_config}. Notably, no per-turn timeout,
file-read limit, or patch-size restriction is imposed---agents may
produce arbitrarily large edits within the task-level budget.
Context window sizes and sampling temperatures follow the default
settings of each model provider's official API; we do not override
these parameters to ensure results reflect production-equivalent
inference behavior.

\paragraph{Scaffold differences.}
OpenCode and Cline differ primarily in their system prompts and tool
invocation patterns. OpenCode uses a structured XML-based action format
with explicit planning steps, while Cline employs a conversational
format with implicit planning. Both expose identical underlying
capabilities (shell execution, file reading/writing, patch submission).
The choice of two scaffolds enables assessment of scaffold-invariant
model capabilities versus scaffold-specific behaviors.

\subsection{Threshold and Hyperparameter Sensitivity}
\label{app:sensitivity}

\paragraph{Coupling threshold $\theta$.}
Table~\ref{tab:theta_sensitivity} reports construction quality metrics
as a function of the coupling threshold $\theta$ on the development set
(10 repositories, 50 gold edges).

\begin{table}[t]
\centering
\caption{Sensitivity to coupling threshold $\theta$ (dev set,
$\Delta{=}90$\,days). \textbf{Bold}: selected configuration.}
\label{tab:theta_sensitivity}
\small
\begin{tabular}{lccc}
\toprule
$\theta$ & \textbf{AP}$\uparrow$ & \textbf{Coverage}$\uparrow$ & \textbf{Exec. Rate}$\uparrow$ \\
\midrule
0.05 & 31.4 & 92.0 & 62.4 \\
0.10 & 35.6 & 88.0 & 71.2 \\
0.15 & 37.6 & 84.0 & 76.8 \\
\textbf{0.20} & 37.2 & 78.0 & 82.4 \\
0.25 & 34.0 & 68.0 & 80.6 \\
0.30 & 30.2 & 58.0 & 78.8 \\
\bottomrule
\end{tabular}
\end{table}

We select $\theta = 0.20$ as it achieves the best tradeoff between
edge precision and executable construction success: AP remains within
0.4\,pp of the peak ($\theta = 0.15$) while the executable
construction rate reaches its maximum. Lower thresholds introduce
spurious edges that increase merge conflicts during rollback
construction, while higher thresholds sacrifice recall of genuine
dependency relationships.

\begin{figure}[t]
\centering
\includegraphics[width=\columnwidth]{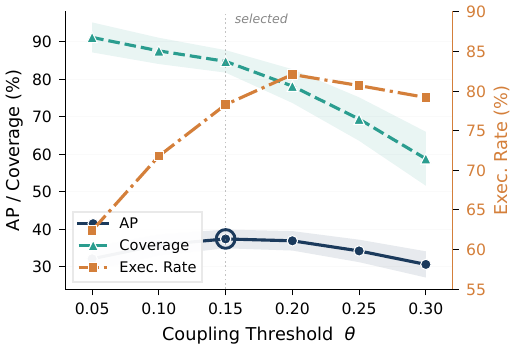}
\caption{Sensitivity of construction quality to coupling threshold
$\theta$. Left axis: AP (solid) and Coverage (dashed); right axis:
executable construction rate. The selected $\theta = 0.20$
(vertical dotted line) balances precision and recall while maintaining
high executability. Shaded region indicates $\pm$1 standard deviation
across development-set repositories.}
\label{fig:theta_sensitivity}
\end{figure}

\paragraph{Search radius $\Delta$.}
Table~\ref{tab:delta_sensitivity} reports sensitivity to the temporal
search radius.

\begin{table}[t]
\centering
\caption{Sensitivity to search radius $\Delta$ (dev set,
$\theta{=}0.20$). \textbf{Bold}: selected configuration.}
\label{tab:delta_sensitivity}
\small
\begin{tabular}{lccc}
\toprule
$\Delta$ (days) & \textbf{AP}$\uparrow$ & \textbf{Coverage}$\uparrow$ & \textbf{Candidates} \\
\midrule
7   & 38.4 & 48.0 & 326 \\
14  & 38.0 & 60.0 & 708 \\
30  & 37.6 & 72.0 & 1{,}482 \\
60  & 37.4 & 77.4 & 3{,}716 \\
\textbf{90}  & 37.2 & 78.0 & 6{,}218 \\
\bottomrule
\end{tabular}
\end{table}

We select $\Delta = 90$ days. The median inter-PR interval in our
corpus is ${\sim}$2--3~weeks, well within this window; 90~days serves
as a conservative upper bound that captures long-lived feature arcs
without omitting genuine dependencies. Coverage saturates by 60~days
and gains only 0.6\,pp from 60 to 90, confirming that
90~days is a safe ceiling rather than a tight fit.

\section{Additional Experimental Results}
\label{app:results}

This section provides extended results for all four research questions.
The dependency graph---as the central computational object---connects
these results: RQ1 validates graph construction quality, RQ2 evaluates
agent performance under graph-conditioned scoring, and RQ3--RQ4
demonstrate that the same graph structure improves both supervised and
reinforcement-based training.

\subsection{RQ1 Extended Results}
\label{app:rq1_extended}

\paragraph{Precision at K curves.}
Table~\ref{tab:p_at_k} reports precision at multiple rank cutoffs for
all methods, providing a more complete view of ranking quality beyond
P@50. Following the main-text protocol, candidate edges from all
30~repositories are merged into a single score-ranked list, and P@K is
the fraction of gold dependency edges among the top~$K$ entries of that
pooled ranking. Because the pooled list is scored globally, P@K is
quantized to multiples of $100/K$ percentage points.

\begin{table}[t]
\centering
\caption{Precision at various rank cutoffs. Candidate edges from all
30~repositories are merged into one score-ranked list; P@K is the
fraction of gold dependency edges among its top~$K$ entries.}
\label{tab:p_at_k}
\setlength{\tabcolsep}{4pt}
\small
\begin{tabular}{lcccc}
\toprule
\textbf{Method} & \textbf{P@10} & \textbf{P@20} & \textbf{P@50} & \textbf{P@100} \\
\midrule
Temporal Window      & 10.0 & 5.0  & 2.0  & 2.0 \\
Traceability Links   & 60.0 & 45.0 & 28.0 & 17.0 \\
File Overlap         & 20.0 & 15.0 & 10.0 & 8.0 \\
Symbol Overlap       & 50.0 & 40.0 & 30.0 & 22.0 \\
\textbf{WEAVE}       & \textbf{70.0} & \textbf{50.0} & \textbf{38.0} & \textbf{29.0} \\
\bottomrule
\end{tabular}
\end{table}

\textsc{WEAVE} attains the highest precision at every cutoff, and no
single baseline is consistently second: Traceability Links is the
strongest alternative at $K\!\leq\!20$, where its explicit cross-PR
references are almost always genuine, but it exhausts its recoverable
edges early (Coverage = 12.8\%) and falls below Symbol Overlap by
$K\!=\!50$. This crossover is precisely why neither signal suffices
alone.

Pooled P@K and per-repository AP (reported in the RQ1 table of the main
text) measure different properties and need not agree. AP is computed
within each repository and then macro-averaged, so it is insensitive to whether
scores are comparable \emph{across} repositories; pooled P@K merges all
repositories into one ranking and therefore additionally rewards
cross-repository score calibration. File Overlap illustrates the gap:
its raw Jaccard scores rank reasonably within a repository (AP 26.6\%)
but are dominated globally by large repositories with many shared
files (P@50 10.0\%). \textsc{WEAVE}'s repository-local inverse-frequency
weighting normalizes locus evidence against each repository's own
modification statistics, which keeps its scores comparable across
repositories and yields close agreement between the two views
(AP 37.7\% vs.\ P@50 38.0\%).

\paragraph{Per-repository AP distributions.}
Across the 30 evaluation repositories, \textsc{WEAVE} achieves:
\begin{itemize}
    \item Median AP: 36.2\% (IQR: 28.4--47.1\%).
    \item AP $>$ 50\% in 6 repositories.
    \item AP $<$ 20\% in 3 repositories (all with
    $<$5 gold edges, limiting statistical power).
\end{itemize}
The next-best method (File Overlap) achieves median AP of
24.8\% (IQR: 18.2--35.6\%).
The macro-average (37.7\%) exceeds the median, reflecting a mild right
skew: a small number of repositories with dense, well-separated
dependency structure achieve high AP, while the bulk of the
distribution is concentrated between the quartiles. The interquartile
ranges of the two methods are disjoint at the median, indicating that
\textsc{WEAVE}'s advantage is distributional rather than driven by a
few favorable repositories.

\paragraph{Locus-type ablation.}
Table~\ref{tab:locus_ablation} evaluates the contribution of each
locus type by training with individual types and with leave-one-out
configurations.

\begin{table}[t]
\centering
\caption{Locus-type ablation. ``Single'': using only that type.
``$-$Type'': removing that type from the full set.}
\label{tab:locus_ablation}
\setlength{\tabcolsep}{4pt}
\small
\begin{tabular}{lcc}
\toprule
\textbf{Configuration} & \textbf{AP}$\uparrow$ & \textbf{Coverage}$\uparrow$ \\
\midrule
\multicolumn{3}{l}{\textit{Single type}} \\
\quad $\mathcal{L}_{\mathrm{ast}}$ only     & 28.4 & 62.1 \\
\quad $\mathcal{L}_{\mathrm{sym}}$ only     & 22.6 & 55.8 \\
\quad $\mathcal{L}_{\mathrm{cfg}}$ only     & 14.2 & 38.4 \\
\quad $\mathcal{L}_{\mathrm{decl}}$ only    & 18.8 & 44.2 \\
\quad $\mathcal{L}_{\mathrm{dep}}$ only     & 12.4 & 32.6 \\
\midrule
\multicolumn{3}{l}{\textit{Leave-one-out}} \\
\quad $-\mathcal{L}_{\mathrm{ast}}$         & 31.2 & 76.8 \\
\quad $-\mathcal{L}_{\mathrm{sym}}$         & 34.6 & 80.2 \\
\quad $-\mathcal{L}_{\mathrm{cfg}}$         & 36.8 & 83.4 \\
\quad $-\mathcal{L}_{\mathrm{decl}}$        & 35.4 & 82.1 \\
\quad $-\mathcal{L}_{\mathrm{dep}}$         & 37.0 & 84.0 \\
\midrule
\textbf{All types (WEAVE)}                  & \textbf{37.7} & \textbf{84.5} \\
\bottomrule
\end{tabular}
\end{table}

$\mathcal{L}_{\mathrm{ast}}$ is the most informative single type
(AP 28.4\%), and its removal causes the largest AP drop
($-$6.5\,pp). However, no single type matches the full combination,
confirming that the five locus types capture complementary evidence
channels. Configuration and dependency loci contribute less
individually but improve coverage by capturing relationships invisible
to code-structural analysis alone.

Two comparisons against the main-text baselines are worth separating.
$\mathcal{L}_{\mathrm{ast}}$ alone (AP 28.4\%) already exceeds the
strongest baseline, File Overlap (26.6\%), showing that moving from
file granularity to declaration granularity is itself the dominant
source of improvement. The Symbol Overlap baseline (24.0\%) is
\emph{not} the counterpart of $\mathcal{L}_{\mathrm{sym}}$ alone
(22.6\%): that baseline matches shared function and class
declarations, a scope spanning both $\mathcal{L}_{\mathrm{ast}}$ and
$\mathcal{L}_{\mathrm{sym}}$, whereas $\mathcal{L}_{\mathrm{sym}}$ in
our decomposition covers only exported symbols, type aliases,
constants, and enum members. The like-for-like comparison is therefore
$\mathcal{L}_{\mathrm{ast}}$ (28.4\%) against Symbol Overlap (24.0\%),
which isolates the contribution of repository-calibrated weighting.

\subsection{RQ3 Full Numerical Results and Token Scaling}
\label{app:rq3_full}

\paragraph{Full SFT comparison table.}
Table~\ref{tab:rq3_full} provides the complete numerical results
underlying the SFT comparison figure of the main text (RQ3).

\begin{table*}[t]
\centering
\caption{Full SFT comparison across data construction strategies
(matched token budget). All values are percentages (\%).
\textbf{Bold}: best per model scale; \underline{underline}: best
overall.}
\label{tab:rq3_full}
\setlength{\tabcolsep}{3.5pt}
\small
\begin{tabular}{llccccc}
\toprule
\textbf{Model} & \textbf{Training Data}
  & \textbf{SWE-V} & \textbf{SWE-Pro} & \textbf{SWE-ML}
  & \textbf{WB-Res} & \textbf{WB-GC} \\
\midrule
\multicolumn{7}{l}{\textit{Qwen3.5-4B}} \\[2pt]
& Base (no SFT)       & 50.4 & 24.2 & 35.3 &  4.0 &  8.6 \\
\cmidrule(l){2-7}
\multicolumn{7}{l}{\quad\footnotesize\textit{External Reference}} \\
& SWE-rebench V2      & 51.8 & 25.4 & 36.6 &  4.7 &  9.5 \\
& daVinci-Agency      & 50.6 & 24.0 & 35.7 &  3.7 &  8.2 \\
\cmidrule(l){2-7}
\multicolumn{7}{l}{\quad\footnotesize\textit{Controlled (same repos, matched tokens)}} \\
& Atomic-PR           & 54.2 & 27.8 & 38.5 &  6.0 & 12.4 \\
& Temporal-Window     & 55.1 & 28.6 & 39.2 &  6.7 & 13.5 \\
& \textbf{WEAVE}
  & \textbf{57.8} & \textbf{31.0}
  & \textbf{41.7} & \textbf{9.0}
  & \textbf{17.8} \\
\midrule
\multicolumn{7}{l}{\textit{Qwen3.5-35B-A3B}} \\[2pt]
& Base (no SFT)       & 69.2 & 44.6 & 60.3 &  7.7 & 14.2 \\
\cmidrule(l){2-7}
\multicolumn{7}{l}{\quad\footnotesize\textit{External Reference}} \\
& SWE-rebench V2      & 70.4 & 45.8 & 61.4 &  8.3 & 15.3 \\
& daVinci-Agency      & 69.4 & 44.2 & 60.5 &  7.3 & 13.8 \\
\cmidrule(l){2-7}
\multicolumn{7}{l}{\quad\footnotesize\textit{Controlled (same repos, matched tokens)}} \\
& Atomic-PR           & 71.8 & 47.6 & 62.7 & 10.3 & 18.6 \\
& Temporal-Window     & 72.4 & 48.3 & 63.2 & 11.0 & 19.7 \\
& \textbf{WEAVE}
  & \underline{\textbf{74.6}} & \underline{\textbf{50.8}}
  & \underline{\textbf{65.9}} & \underline{\textbf{14.0}}
  & \underline{\textbf{24.8}} \\
\bottomrule
\end{tabular}
\end{table*}

\begin{figure}[t]
\centering
\includegraphics[width=\columnwidth]{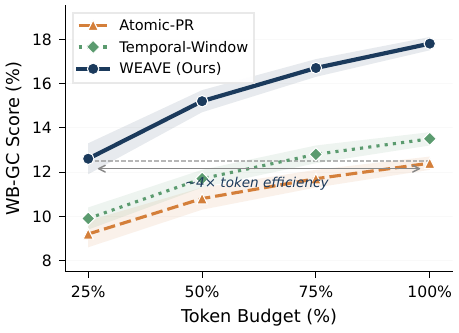}
\caption{Token-scaling curves for SFT training (Qwen3.5-4B, measured
by WEAVE-Bench Graph Credit). \textsc{WEAVE} (solid blue) achieves at
25\% of the token budget what Atomic-PR (dashed orange) achieves at
100\%, indicating a token-efficiency advantage of at least
$\sim$4$\times$. Temporal-Window (dotted green) falls between the two.
The three curves differ in slope rather than offset, indicating that
dependency-structured supervision is denser per token rather than
merely better initialized. Error bars show $\pm$1 std across 3 random
seeds.}
\label{fig:token_scaling}
\end{figure}

\paragraph{Token-scaling analysis.}
We evaluate how each data construction strategy scales with training
token budget by training Qwen3.5-4B at 25\%, 50\%, 75\%, and 100\% of
the full budget (${\sim}$158\,M tokens; \S\ref{app:trajectory}). Table~\ref{tab:token_scaling}
reports the resulting WB-GC scores and Figure~\ref{fig:token_scaling}
plots the corresponding curves.

\begin{table}[t]
\centering
\caption{Token-scaling analysis (Qwen3.5-4B, WB-GC metric).}
\label{tab:token_scaling}
\setlength{\tabcolsep}{4pt}
\small
\begin{tabular}{lcccc}
\toprule
\textbf{Budget} & \textbf{25\%} & \textbf{50\%} & \textbf{75\%} & \textbf{100\%} \\
\midrule
Atomic-PR       & 9.2  & 10.8 & 11.7 & 12.4 \\
Temporal-Window & 9.9  & 11.7 & 12.8 & 13.5 \\
\textbf{WEAVE}  & \textbf{12.6} & \textbf{15.2} & \textbf{16.7} & \textbf{17.8} \\
\bottomrule
\end{tabular}
\end{table}

All three strategies scale approximately log-linearly in token budget,
gaining a roughly constant amount per doubling
($\sim$1.6\,pp for Atomic-PR, $\sim$1.8\,pp
for Temporal-Window, $\sim$2.6\,pp for \textsc{WEAVE}),
with no strategy saturating within the budgets tested. The curves are
separated by slope rather than by offset, which is the signature of a
difference in supervision density rather than a fixed head start.
\textsc{WEAVE} reaches at 25\% budget (12.6\%) what Atomic-PR reaches
only at 100\% (12.4\%). Since we do not evaluate below 25\%, this
establishes a token-efficiency advantage of \emph{at least}
$\sim$4$\times$ at this operating point rather than an exact ratio.

\subsection{RQ4 Component Ablation and Curriculum}
\label{app:rq4_ablation}

\paragraph{Reward component ablation.}
Table~\ref{tab:reward_ablation} evaluates the contribution of each
component in $R_{\mathrm{WEAVE}}$ by progressive removal.

\begin{table}[t]
\centering
\caption{Reward component ablation (Qwen3.5-4B, after RL). All runs
start from the same \textsc{WEAVE}-SFT checkpoint with matched rollout
budgets. $\Delta$ is the WB-Res difference relative to full
$R_{\mathrm{WEAVE}}$. After each removal the remaining coefficients are
renormalized to sum to one, so differences reflect the missing signal
rather than a change in reward scale.}
\label{tab:reward_ablation}
\setlength{\tabcolsep}{3.5pt}
\small
\begin{tabular}{lcccc}
\toprule
\textbf{Reward Config} & \textbf{WB-Res} & \textbf{WB-GC} & \textbf{SWE-V} & $\Delta$ \\
\midrule
\textit{SFT only (no RL)}
  & \textit{9.0} & \textit{17.8} & \textit{57.8} & \textit{$-$3.7} \\
\midrule
$R_{\mathrm{WEAVE}}$ (full)
  & 12.7 & 23.4 & 61.0 & -- \\
$-$ GraphCredit
  & 11.0 & 19.0 & 59.4 & $-$1.7 \\
$-$ Resolved
  & 11.6 & 22.3 & 60.1 & $-$1.1 \\
$-$ FlatF2P
  & 12.1 & 22.7 & 60.5 & $-$0.6 \\
$-$ Reg$_{\mathrm{P2P}}$
  & 12.4 & 23.1 & 60.7 & $-$0.3 \\
\midrule
Binary only (Resolved)
  & 9.7  & 18.4 & 58.3 & $-$3.0 \\
Flat F2P only
  & 10.7 & 18.8 & 59.0 & $-$2.0 \\
\bottomrule
\end{tabular}
\end{table}

Graph Credit is the most impactful individual component: its removal
reduces WB-GC by 4.4\,pp and WB-Res by
1.7\,pp. The Resolved bonus is the second most important
component, providing an explicit signal for full-task completion. The
regression penalty ($\operatorname{Reg}_{\mathrm{P2P}}$) has minimal
impact on performance metrics but is critical for maintaining code
quality: among the four reward variants compared in RQ4 the P2P
regression rate stays at 1.8\%, whereas removing the
penalty from $R_{\mathrm{WEAVE}}$ raises it to 4.7\%.
This ablation-only configuration is the sole setting that exceeds the
2.3\% regression ceiling reported in the main text.

The two single-signal rewards bound the ablation from below. Binary
\textsc{Resolved} improves over the SFT checkpoint by at most
0.7\,pp on any metric: on \textsc{WEAVE-Bench} a rollout
earns reward only when every constituent PR is solved, so most rollouts
return identical zero reward and GRPO receives almost no
advantage signal to differentiate them. Flat F2P restores a dense
signal and recovers roughly half the gap to $R_{\mathrm{WEAVE}}$, but
because it credits downstream test passes irrespective of upstream
state, it optimizes precisely the phantom credit that Graph Credit is
designed to remove. Removing Graph Credit from the full reward
(retaining Resolved, FlatF2P, and the regression penalty) remains
better than Flat F2P alone, confirming that the residual components
contribute independently.

\paragraph{DAG-complexity curriculum.}
Table~\ref{tab:curriculum} compares the three-phase curriculum against
flat (uniform random) task scheduling.

\begin{table}[t]
\centering
\caption{Curriculum scheduling comparison (Qwen3.5-4B). Phases
defined by DAG complexity: P1 (depth $\leq$2, $|V|\leq$3),
P2 (depth $\leq$3, $|V|\leq$5), P3 (all tasks).}
\label{tab:curriculum}
\setlength{\tabcolsep}{4pt}
\small
\begin{tabular}{lccc}
\toprule
\textbf{Scheduling} & \textbf{WB-Res} & \textbf{WB-GC} & \textbf{SWE-V} \\
\midrule
Flat (uniform random)   & 11.7 & 22.2 & 60.2 \\
\textbf{3-phase curriculum} & \textbf{12.7} & \textbf{23.4} & \textbf{61.0} \\
\midrule
$\Delta$ & +1.0\,pp & +1.2\,pp & +0.8\,pp \\
\bottomrule
\end{tabular}
\end{table}

The three-phase curriculum provides consistent improvements of
$\sim$1.0\,pp across metrics. We observe that this benefit arises
primarily in the first 30\% of training: the curriculum prevents early
divergence on overly complex tasks before the policy has developed basic
multi-file editing capabilities.

\paragraph{Curriculum phase details.}
The three phases are defined by DAG complexity thresholds and are
allocated equal thirds of the update budget. Task counts and the
schedule itself are given in \S\ref{app:rl}; here we report the
difficulty gradient the curriculum induces:
\begin{itemize}
    \item \textbf{Phase 1} (0--33\% of updates): depth $\leq$2 and
    $|V| \leq 3$. Mean $|V|$: 2.39. Mean gold patch: 166 lines.
    \item \textbf{Phase 2} (33--66\%): depth $\leq$3 and $|V| \leq 5$.
    Mean $|V|$: 3.03. Mean gold patch: 211 lines.
    \item \textbf{Phase 3} (66--100\%): all tasks, sampled uniformly.
    Mean $|V|$: 4.16. Mean gold patch: 289 lines.
\end{itemize}
Mean patch size scales near-linearly with node count at roughly
70~lines per PR node, so the curriculum increases edit scope by
${\sim}$1.7$\times$ from the first phase to the last.

\section{Training Details}
\label{app:training}

The same dependency graph that defines evaluation tasks and
Graph Credit scoring also structures the training pipeline. This
section details how \textsc{WEAVE} leverages the PR dependency DAG
to generate high-quality SFT trajectories and graph-structured RL
rewards, completing the unified framework from construction through
optimization.

\subsection{Trajectory Generation and Quality Filtering}
\label{app:trajectory}

\paragraph{Trajectory generation setup.}
Training trajectories are generated using the Cline scaffold with the
following configuration:
\begin{itemize}
    \item \textbf{Generator model}: GLM-5.2.
    \item \textbf{Max turns}: 200.
    \item \textbf{Context window}: 128K tokens.
    \item \textbf{Temperature}: 0.6 (for diversity).
    \item \textbf{Tasks attempted}: 3{,}200
    (all of \textsc{WEAVE-Train}).
    \item \textbf{Attempts per task}: 3
    (best-of-3 by Graph Credit).
\end{itemize}

\paragraph{Choice of generator.}
We generate with GLM-5.2 rather than the strongest evaluated
configuration (Claude Opus~4.7 + Cline, per the RQ2 results in the main
text) for three reasons. First, GLM-5.2 has openly released weights, so the
resulting corpus can be published without the derived-data restrictions
that proprietary model outputs carry, and the generation stage is
self-hostable and therefore independently reproducible. Second, GLM-5.2
is the second-strongest configuration on \textsc{WEAVE-Bench}
(43.0\% versus 48.7\% Resolved); the best-of-3 protocol absorbs most of
this 5.7\,pp gap at the corpus level. Third, generator and student share
a broadly similar tool-invocation format, which reduces format
adaptation during distillation. The cost is throughput: GLM-5.2 averages
138.2 turns and 7.52\,M cumulative tokens per benchmark task, against
46.8 turns and 2.11\,M for Claude Opus~4.7. This 3.6$\times$ generation
overhead is why we cap attempts at three per task rather than sampling
more widely.

\paragraph{Quality filtering.}
Generated trajectories pass through five filters. These are deliberately
strict: a trajectory that satisfies downstream tests while leaving
upstream prerequisites unimplemented exhibits precisely the phantom
credit that \textsc{WEAVE-Bench} is designed to expose, and distilling
such behavior into the student would be self-defeating. We therefore
gate on Graph Credit rather than on flat test pass rate.

\begin{enumerate}
    \item \textbf{Dependency-consistent success filter}: Retain
    trajectories satisfying both
    $\operatorname{GraphCredit}(G_A)\geq 0.85$ and
    $\min_{v\in V_A} c_v \geq 0.50$. The first condition requires
    near-complete resolution of the task graph. The second is applied
    per node and forbids any single node from remaining grossly
    unimplemented; because $c_v$ is already bottlenecked by its
    ancestors under the recursive definition of Graph Credit in the
    main text, it constrains every dependency path simultaneously. A trajectory with high Flat-F2P
    but an unresolved upstream node fails this filter by construction.

    \item \textbf{Length filter}: Discard trajectories shorter than
    8~turns, and any trajectory that terminated by exhausting the
    200-turn or context budget rather than by submitting a patch, since
    truncated rollouts are incomplete by construction. The resulting
    effective upper bound is 150~turns, 75\% of the interaction cap.

    \item \textbf{Regression filter}: Discard trajectories exhibiting
    any P2P regression. Failing P2P tests are re-executed three times
    and counted as regressions only when they fail consistently, so
    that flaky tests do not discard otherwise valid trajectories.

    \item \textbf{Degenerate pattern filter}: Remove trajectories with
    three or more consecutive identical tool calls (same tool, same
    arguments), or with edit churn above 0.5---the fraction of edit
    operations subsequently overwritten or reverted within the same
    trajectory---which indicates thrashing rather than incremental
    progress.

    \item \textbf{Leakage filter}: Discard trajectories in which the
    agent recovered the original change through repository history or
    network access, detected by scanning tool-call arguments for
    \texttt{git log}, \texttt{git show}, \texttt{.git} traversal, and
    outbound requests. Task instantiation already removes recoverable
    target commits and repository history; this filter enforces the
    same guarantee at the trajectory level, where an agent could
    otherwise reach them at run time.
\end{enumerate}

\noindent Surviving rollouts are deduplicated per task: we keep the
attempt with the highest Graph Credit, break ties by shorter trajectory
length, and treat remaining attempts whose final patches exceed 0.9
normalized edit-distance similarity to the retained one as duplicates.

\paragraph{Corpus statistics after filtering.}
Table~\ref{tab:traj_funnel} reports the yield at each stage.

\begin{table}[t]
\centering
\caption{Trajectory generation and filtering yield. All rows count
rollouts except the last, which counts one retained trajectory per task.}
\label{tab:traj_funnel}
\setlength{\tabcolsep}{5pt}
\small
\begin{tabular}{lcc}
\toprule
\textbf{Stage} & \textbf{Retained} & \textbf{Pass rate} \\
\midrule
Rollouts generated (3{,}200\,$\times$\,3) & 9{,}600 & -- \\
\midrule
Dependency-consistent success & 4{,}320 & 45.0\% \\
Length                        & 4{,}058 & 93.9\% \\
Regression                    & 3{,}857 & 95.0\% \\
Degenerate pattern            & 3{,}699 & 95.9\% \\
Leakage                       & 3{,}644 & 98.5\% \\
\midrule
\textbf{Per-task deduplication} & \textbf{2{,}304} & \textbf{72.0\%}$^{\dagger}$ \\
\bottomrule
\end{tabular}

\vspace{2pt}
{\footnotesize $^{\dagger}$Fraction of the 3{,}200 tasks retaining at
least one trajectory.}
\end{table}

The final corpus contains 2{,}304 trajectories covering 72.0\% of
\textsc{WEAVE-Train} tasks, with mean length 61.4~turns and mean size
68.4K~tokens, totaling ${\sim}$158\,M training tokens. Task coverage
remains approximately uniform across dependency depth---74.1\% for
tasks of depth\,$\leq$\,2 versus 69.4\% for depth\,$\geq$\,4---so the
strict success filter does not bias the corpus toward shallow
structures, which would otherwise undercut the supervision signal that
motivates dependency-structured construction.

\subsection{SFT Configuration}
\label{app:sft}

Table~\ref{tab:sft_config} details the SFT hyperparameters for both
model scales. We report only parameters that affect the optimization
objective or its trajectory; sharding strategy, device placement, and
accumulation factors are implementation choices that follow from the
global batch size and are omitted.

\begin{table}[t]
\centering
\caption{SFT training configuration. Search ranges are given for
parameters tuned on the held-out calibration set; ``--'' marks values
fixed a priori.}
\label{tab:sft_config}
\setlength{\tabcolsep}{3.5pt}
\small
\begin{tabular}{lccc}
\toprule
\textbf{Parameter} & \textbf{4B} & \textbf{35B-A3B} & \textbf{Search range} \\
\midrule
Learning rate        & 5e-6 & 3e-6 & \{2,3,5,10\}e-6 \\
LR schedule          & \multicolumn{2}{c}{Cosine, 10\% floor} & -- \\
Warmup ratio         & 0.03 & 0.03 & \{0.01,\,0.03,\,0.05\} \\
Global batch (seq.)  & 16   & 16   & -- \\
Training epochs      & 3    & 3    & \{2,\,3,\,4\} \\
Max sequence length  & \multicolumn{2}{c}{131{,}072} & -- \\
Weight decay         & 0.01 & 0.01 & -- \\
Adam $\beta_1$/$\beta_2$ & \multicolumn{2}{c}{0.9\,/\,0.95} & -- \\
Gradient clipping    & 1.0  & 1.0  & -- \\
Precision            & \multicolumn{2}{c}{BF16} & -- \\
Loss masking         & \multicolumn{2}{c}{Assistant tokens only} & -- \\
\bottomrule
\end{tabular}
\end{table}

Three choices warrant comment. \textbf{Loss masking}: gradients are
computed only on assistant-generated tokens; tool results, file
contents, and test output are supplied as context but excluded from the
objective, since these constitute environment observations rather than
target behavior. \textbf{Sequence length}: the 131{,}072-token window
matches the generator's context budget, so no retained trajectory is
truncated during training; training at this length uses sequence
parallelism. \textbf{Learning rate}: with only 2{,}304 trajectories, the
corpus is small enough that the higher end of the search range
overfits within one epoch. Qwen3.5-35B-A3B activates roughly 3B
parameters per token, so its optimal rate is close to---though slightly
below---that of the 4B dense model rather than an order of magnitude
lower.

Both scales use identical batch size and epoch count so that the
controlled comparison in RQ3 varies only task construction. The
resulting schedule is 144 optimizer steps per epoch and 432 steps
overall, of which the first 13 are warmup. All runs use a single fixed
random seed; reported results are single-run unless stated otherwise.

\paragraph{Data mixing strategy.}
For controlled comparisons, we ensure matched token budgets across all
data sources:
\begin{itemize}
    \item Each training arm uses ${\sim}$158\,M tokens.
    \item For external references (SWE-rebench V2, daVinci-Agency),
    we sample trajectories uniformly to match the token budget.
    \item For controlled variants (Atomic-PR, Temporal-Window, WEAVE),
    trajectories are generated from the same repository pool using the
    same generator and quality filter, differing only in task
    construction.
    \item All variants use identical SFT configuration
    (Table~\ref{tab:sft_config}).
\end{itemize}

\paragraph{Checkpoint selection.}
We select the final checkpoint by Graph Credit on a held-out calibration
set of 50 tasks disjoint from \textsc{WEAVE-Bench}, evaluating every
36~steps (one quarter epoch) for 12 evaluation points across the
432-step schedule.

\subsection{RL Configuration}
\label{app:rl}

Table~\ref{tab:rl_config} details the GRPO
configuration for RQ4. Unlike SFT, RL requires no gold trajectory---only
an executable environment---so the prompt pool is the full
\textsc{WEAVE-Train} set of 3{,}200 tasks rather than the 2{,}304 tasks
for which a demonstration survived filtering.

We specify the interaction budget in terms of the quantities that
define it---pool size, group size, tasks per update, and passes over the
pool---and report the induced update and episode counts as derived
values, so that the budget can be reconstructed exactly.

\begin{table}[t]
\centering
\caption{RL (GRPO) training configuration. Rows marked $\ast$ are
derived from the rows above them. Search ranges are given for parameters
tuned on the calibration set.}
\label{tab:rl_config}
\setlength{\tabcolsep}{3.5pt}
\small
\begin{tabular}{lccc}
\toprule
\textbf{Parameter} & \textbf{4B} & \textbf{35B-A3B} & \textbf{Search range} \\
\midrule
RL prompt pool          & \multicolumn{2}{c}{3{,}200 tasks} & -- \\
Rollouts per task ($G$) & 8 & 8 & -- \\
Tasks per update        & 16 & 16 & -- \\
Passes over pool        & 2 & 2 & -- \\
$\ast$ Policy updates   & 400 & 400 & -- \\
$\ast$ Total episodes   & \multicolumn{2}{c}{51{,}200} & -- \\
\midrule
Learning rate           & 1e-6 & 5e-7 & \{5e-7,\,1e-6,\,2e-6\} \\
LR schedule             & \multicolumn{2}{c}{Cosine decay} & -- \\
KL coefficient ($\beta$) & 0.01 & 0.01 & \{0.005,\,0.01,\,0.04\} \\
Clip ratio ($\epsilon$) & 0.2 & 0.2 & -- \\
Rollout temperature     & 1.0 & 1.0 & \{0.8,\,1.0\} \\
Max turns per episode   & 200 & 200 & -- \\
Reference policy        & \multicolumn{2}{c}{SFT checkpoint} & -- \\
Advantage               & \multicolumn{2}{c}{Group-relative} & -- \\
Precision               & \multicolumn{2}{c}{BF16} & -- \\
\bottomrule
\end{tabular}
\end{table}

Both scales share the same interaction budget so that the reward
comparison in RQ4 is not confounded by differing amounts of experience.
The rollout temperature of 1.0 is deliberate: GRPO derives its advantage
signal from variation \emph{within} each group of $G$ rollouts, so
insufficient sampling diversity collapses the group to a common reward
and contributes no gradient.

\paragraph{Reward coefficients.}
The reward $R_{\mathrm{WEAVE}}$ uses the following coefficients
(selected via grid search on calibration set):
\begin{itemize}
    \item $\lambda_{\mathrm{graph}} = 0.5$ (dominant component)
    \item $\lambda_{\mathrm{solve}} = 0.25$ (full-task bonus)
    \item $\lambda_{\mathrm{f2p}} = 0.15$ (flat progress)
    \item $\lambda_{\mathrm{reg}} = 0.1$ (regression penalty)
\end{itemize}

\paragraph{Advantage normalization.}
GRPO is critic-free: for each task we sample $G=8$ rollouts, score each
with $R_{\mathrm{WEAVE}}$, and take the advantage of a rollout to be its
reward centered and scaled by the mean and standard deviation
\emph{within its own group}. No value network, and hence no generalized
advantage estimation, is used. Group-relative normalization makes the
gradient invariant to task difficulty, so that easy and hard tasks
contribute comparably.

Group-relative normalization also explains why reward density matters
here beyond the usual sparse-reward argument: a group whose $G$ rollouts
all receive the same reward contributes no gradient at all, so a sparse
reward wastes rollout budget outright rather than merely learning
slowly. Because Graph Credit is continuous and dependency-conditioned,
its groups almost always exhibit within-group spread. The empirical
consequence is quantified in \S\ref{app:rq4_ablation}.

\paragraph{Three-phase curriculum.}
RL training follows a DAG-complexity curriculum over the 3{,}200-task
pool. Phases are cumulative---each admits all tasks from the previous
phase---and are allocated equal thirds of the 400 policy updates:
\begin{itemize}
    \item \textbf{Phase 1} (updates 1--132): DAGs with depth $\leq$2
    and $|V| \leq 3$. Eligible tasks: 1{,}048.
    \item \textbf{Phase 2} (updates 133--264): DAGs with depth $\leq$3
    and $|V| \leq 5$. Eligible tasks: 1{,}864.
    \item \textbf{Phase 3} (updates 265--400): all tasks, sampled
    uniformly. Eligible tasks: 3{,}200.
\end{itemize}
Phase membership follows directly from the \textsc{WEAVE-Train} node and
depth distributions (Table~\ref{tab:train_nodes}); depth is measured as
the number of edges on the longest path, consistent with
\S\ref{app:depth}.

\section{Dataset Statistics and Case Studies}
\label{app:dataset}

This section provides comprehensive statistics on \textsc{WEAVE-Bench}
and \textsc{WEAVE-Train}, followed by qualitative case studies
demonstrating how Graph Credit operates on concrete dependency
structures produced by the construction pipeline.

\subsection{WEAVE-Bench and WEAVE-Train Statistics}
\label{app:stats}

\paragraph{Language distribution.}
Table~\ref{tab:lang_dist} reports the language composition of both
\textsc{WEAVE-Bench} and \textsc{WEAVE-Train}.

\begin{table}[t]
\centering
\caption{Language distribution across dataset splits.}
\label{tab:lang_dist}
\small
\begin{tabular}{lcc}
\toprule
\textbf{Language} & \textbf{WEAVE-Bench} & \textbf{WEAVE-Train} \\
\midrule
Python      & 50 (16.7\%)  & 720 (22.5\%) \\
TypeScript  & 50 (16.7\%)  & 680 (21.3\%) \\
JavaScript  & 50 (16.7\%)  & 640 (20.0\%) \\
Go          & 50 (16.7\%)  & 480 (15.0\%) \\
Kotlin      & 50 (16.7\%)  & 360 (11.3\%) \\
Java        & 50 (16.7\%)  & 320 (10.0\%) \\
\midrule
\textbf{Total} & \textbf{300} & \textbf{3{,}200} \\
\bottomrule
\end{tabular}
\end{table}

\paragraph{DAG structure statistics.}
Table~\ref{tab:dag_stats} characterizes the dependency graph structures.

\begin{table}[t]
\centering
\caption{DAG structural statistics across splits. Depth counts
nodes on the longest path, matching the benchmark distribution in
Figure~\ref{fig:weave_bench_statistics}. Training depth is converted from
edge count by adding one; curriculum thresholds use edge count.}
\label{tab:dag_stats}
\setlength{\tabcolsep}{3.5pt}
\small
\begin{tabular}{lcccc}
\toprule
\textbf{Metric} & \multicolumn{2}{c}{\textbf{WEAVE-Bench}} & \multicolumn{2}{c}{\textbf{WEAVE-Train}} \\
\cmidrule(lr){2-3} \cmidrule(lr){4-5}
& Mean & Median & Mean & Median \\
\midrule
Nodes ($|V|$)         & 4.91 & 5   & 4.16 & 4 \\
Edges ($|E|$)         & 4.99 & 5   & 4.28 & 4 \\
Depth (path nodes)    & 3.93 & 4   & 3.71 & 4 \\
Width (max anti-chain) & 1.50 & 1  & 1.31 & 1 \\
\bottomrule
\end{tabular}
\end{table}

Table~\ref{tab:train_nodes} gives the full node-count distribution for
\textsc{WEAVE-Train}, which determines the eligible task pool at each
phase of the RL curriculum (\S\ref{app:rl}).

\begin{table}[t]
\centering
\caption{Node-count distribution of \textsc{WEAVE-Train}
($|V|$ mean 4.16, median 4).}
\label{tab:train_nodes}
\setlength{\tabcolsep}{4pt}
\small
\begin{tabular}{lccccccc}
\toprule
$|V|$ & 2 & 3 & 4 & 5 & 6 & 7 & 8 \\
\midrule
Tasks & 640 & 704 & 640 & 480 & 352 & 224 & 160 \\
Share & 20.0 & 22.0 & 20.0 & 15.0 & 11.0 & 7.0 & 5.0 \\
\bottomrule
\end{tabular}
\end{table}

\paragraph{Topology distribution.}
\begin{itemize}
    \item \textbf{Linear chains} (all nodes have in/out-degree
    $\leq$1): 44.3\% of \textsc{WEAVE-Bench} tasks.
    \item \textbf{Diamond structures} (fork followed by join):
    15.7\%.
    \item \textbf{General DAGs} (mixed fork/join patterns):
    15.0\%.
    \item \textbf{Fan structures} (one root, multiple leaves):
    13.3\%.
    \item \textbf{Cross structures} (multiple interleaved
    dependencies): 11.7\%.
\end{itemize}

\paragraph{Patch and test statistics.}

\begin{table}[t]
\centering
\caption{Patch and test statistics for WEAVE-Bench.}
\label{tab:patch_stats}
\small
\begin{tabular}{lcccc}
\toprule
\textbf{Metric} & \textbf{Mean} & \textbf{Median} & \textbf{P25} & \textbf{P75} \\
\midrule
Gold patch (lines)    & 378  & 274  & 139 & 528 \\
Files modified        & 8.4  & 6    & 3   & 12 \\
F2P tests             & 18.6 & 12   & 6   & 24 \\
P2P tests             & 142  & 87   & 34  & 198 \\
Attributed tests (\%) & 72.8 & 76.4 & 61.2& 86.3 \\
\bottomrule
\end{tabular}
\end{table}

\paragraph{Repository coverage.}
\textsc{WEAVE-Bench} draws from 52 unique repositories. The
most-represented repository contributes 23 tasks (7.7\%); the top-12
repositories account for $\sim$60\% of tasks, with 12
repositories contributing only a single task each.
\textsc{WEAVE-Train} covers $\sim$120 repositories.

\subsection{Qualitative Examples}
\label{app:cases}

We present two case studies illustrating how Graph Credit reveals
phantom credit in specific tasks.

\begin{figure*}[t]
\centering
\includegraphics[width=\textwidth]{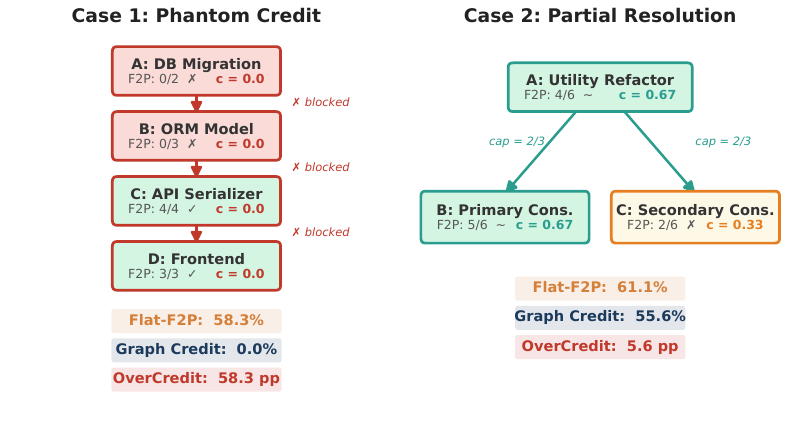}
\caption{Dependency graphs for Case Studies 1 and 2. \textbf{Left}:
4-node DAG (Case~1). Node~A (DB migration) $\to$ Node~B (ORM model)
$\to$ Node~C (API serializer) $\to$ Node~D (frontend). Green fill
indicates agent-completed nodes; red fill indicates unresolved.
\textbf{Right}: 3-node DAG (Case~2). Node~A (utility refactor)
is partially resolved (4/6 tests) and branches to Node~B
(primary consumer, 5/6 tests) and Node~C (secondary consumer, 2/6 tests).
Effective credits are $c_A=c_B=2/3$ and $c_C=1/3$; Node~B is capped
by its upstream prerequisite. Right-panel fill intensity encodes $c_v$.}
\label{fig:case_dags}
\end{figure*}

\paragraph{Case 1: Database migration with downstream API changes.}

\textbf{Task structure:} A 4-node DAG in a Python web application.
Node~A introduces a new database column; Node~B modifies the ORM model
to expose the column; Node~C updates the API serializer; Node~D adds
frontend validation. The dependency graph captures that C and D cannot
function correctly without B's ORM changes, which in turn require A's
schema migration.

\textbf{Agent behavior (Claude Opus 4.7 / Cline):} The agent
successfully implements Nodes~C and~D (API serializer and frontend
validation) by pattern-matching existing code, but fails to modify the
database migration (Node~A) or ORM model (Node~B). The downstream
tests for Nodes~C and~D pass because the test fixtures mock the
database layer.

\textbf{Key F2P tests and attribution:}
\begin{itemize}
    \item Node~A: \texttt{test\_migration\_adds\_column},
    \texttt{test\_rollback\_migration} (2 tests, both fail)
    \item Node~B: \texttt{test\_model\_field\_exists},
    \texttt{test\_model\_serialization},
    \texttt{test\_model\_validation} (3 tests, all fail)
    \item Node~C: \texttt{test\_api\_serialize\_new\_field},
    \texttt{test\_api\_response\_schema},
    \texttt{test\_api\_pagination\_with\_field},
    \texttt{test\_api\_filter\_by\_field} (4 tests, all pass via mock)
    \item Node~D: \texttt{test\_frontend\_validation\_rule},
    \texttt{test\_form\_field\_rendering},
    \texttt{test\_error\_message\_display} (3 tests, all pass via mock)
\end{itemize}

\textbf{Scoring comparison:}
\begin{itemize}
    \item \textbf{Flat-F2P}: 58.3\% (7/12 attributed tests pass).
    \item \textbf{Graph Credit}: 0.0\% (Nodes~C and~D depend on
    Node~B, which depends on Node~A; since $c_A = 0$, all
    downstream credits are capped at 0).
    \item \textbf{OverCredit}: 58.3\,pp.
\end{itemize}

This case demonstrates the most extreme form of phantom credit: the
agent receives majority credit under flat scoring despite implementing
none of the foundational changes. Graph Credit correctly identifies
that no genuine progress has been made toward a coherent solution.

\paragraph{Case 2: Partial resolution with mixed upstream completion.}

\textbf{Task structure:} A 3-node DAG in a TypeScript project.
Node~A refactors a shared utility module; Node~B (depends on~A) updates
the primary consumer; Node~C (depends on~A) updates the secondary
consumer. This fork structure tests whether Graph Credit correctly
attributes partial completion across parallel downstream nodes that
share a common dependency.

\textbf{Agent behavior (GLM-5.2 / OpenCode):} The agent refactors
Node~A but misses two behaviors of the original utility (edge-case
handling and error propagation), implements Node~B (primary consumer
update) against the new interface, and produces an incorrect
implementation of Node~C that passes only 2 of 6 attributed tests.

\textbf{Key F2P tests and attribution:}
\begin{itemize}
    \item Node~A: \texttt{test\_util\_new\_api},
    \texttt{test\_util\_backward\_compat},
    \texttt{test\_util\_edge\_cases},
    \texttt{test\_util\_error\_handling},
    \texttt{test\_util\_types},
    \texttt{test\_util\_async} (6 tests, 4 pass;
    \texttt{edge\_cases} and \texttt{error\_handling} fail)
    \item Node~B: \path{test_primary_consumer_integration},
    \texttt{test\_primary\_consumer\_error\_path},
    \texttt{test\_primary\_consumer\_caching},
    \texttt{test\_primary\_consumer\_retry},
    \texttt{test\_primary\_consumer\_events},
    \texttt{test\_primary\_consumer\_metrics} (6 tests, 5 pass)
    \item Node~C: \texttt{test\_secondary\_consumer\_basic},
    \texttt{test\_secondary\_consumer\_batch},
    \texttt{test\_secondary\_consumer\_timeout},
    \texttt{test\_secondary\_consumer\_fallback},
    \texttt{test\_secondary\_consumer\_logging},
    \texttt{test\_secondary\_consumer\_config} (6 tests, 2 pass)
\end{itemize}

\textbf{Scoring comparison:}
\begin{itemize}
    \item \textbf{Flat-F2P}: 61.1\% (11/18 attributed tests pass:
    4/6 for A, 5/6 for B, 2/6 for C).
    \item \textbf{Graph Credit}: 55.6\%. Node credits:
    $c_A = 4/6 = 0.67$ (no parents),
    $c_B = \min(5/6,\, c_A) = 0.67$,
    $c_C = \min(2/6,\, c_A) = 0.33$. Mean
    $= (2/3 + 2/3 + 1/3)/3 = 5/9\approx 0.556$.
    \item \textbf{OverCredit}: 5.6\,pp.
\end{itemize}

This fork exercises both branches of the recursive minimum. Node~B's
local score (5/6) exceeds its parent's completion ($c_A = 2/3$) and is
therefore capped: flat scoring would report the primary consumer as
nearly complete even though two behaviors of the utility module it
consumes are still missing. Node~C, by contrast, already scores below
its parent, so its own local completion remains the binding constraint
and Graph Credit leaves it unchanged. The entire 5.6\,pp OverCredit is
thus attributable to Node~B---precisely the structurally unsupported
portion of its score.

\paragraph{Aggregate patterns.}
Across all 4,800 agent runs on \textsc{WEAVE-Bench}:
\begin{itemize}
    \item 68\% of runs exhibit $>$0\,pp OverCredit.
    \item 23\% exhibit $>$15\,pp OverCredit (substantial
    phantom credit).
    \item 8\% exhibit $>$30\,pp OverCredit (severe
    misrepresentation under flat scoring).
    \item The correlation between Flat-F2P and actual Resolved is
    $\rho = 0.69$, while Graph Credit achieves
    $\rho = 0.82$---a 19\% improvement
    in diagnostic fidelity.
\end{itemize}

\end{document}